\documentclass{amsart}[10pt]

\usepackage{amscd,amssymb,latexsym,url,verbatim,graphicx,color}
\usepackage{tikz,tikz-cd}
\usetikzlibrary{decorations.pathmorphing}

\usepackage{cases,amsmath}

\usepackage{tikz,tikz-cd}

\usepackage[dvips]{epsfig}

\title[Protocol complexes for immediate snapshot r/w model]
{Standard protocol complexes for the immediate snapshot
read/write model}

\author{Dmitry N. Kozlov}

\address{Department of Mathematics, University of Bremen, 28334
  Bremen, Federal Republic of Germany}

\email{dfk@math.uni-bremen.de}

\newtheorem{theorem}{Theorem}[section]
\newtheorem{df}[theorem]{Definition}
\newtheorem{thm}[theorem]{Theorem} \newtheorem{lemma}[theorem]{Lemma}
\newtheorem{prop}[theorem]{Proposition}
\newtheorem{lm}[theorem]{Lemma}
\newtheorem{crl}[theorem]{Corollary}

\newtheorem{rem}[theorem]{Remark}
\newtheorem{conj}[theorem]{Conjecture} \newcommand{\nin}{\noindent}
\newcommand{\pr}{\nin{\bf Proof.} }

\newcommand{\act}{\text{\rm act}\,}

\newcommand{\ck}{{\mathcal K}}

\newcommand{\cn}{{\mathcal N}}

\newcommand{\cs}{{\mathcal S}}
\newcommand{\codim}{\text{\rm codim}\,}
\newcommand{\csn}{\cs_{\zz_+}}
\newcommand{\da}{\Delta}
\newcommand{\dar}{\downarrow}

\newcommand{\es}{\emptyset}

\newcommand{\hra}{\hookrightarrow}
\newcommand{\inte}{\text{\rm int}\,}

\newcommand{\last}{\text{\rm last}\,}

\newcommand{\mqed}{\quad\Box}

\newcommand{\pass}{\text{\rm pass}\,}
\newcommand{\pnt}{round counter }
\newcommand{\pnts}{round counters }
\newcommand{\ra}{\rightarrow}

\newcommand{\st}{\text{\rm{st}}}
\newcommand{\sm}{\setminus}
\newcommand{\smax}{\text{\rm smax}\,}

\newcommand{\supp}{\text{\rm supp}\,}

\newcommand{\tr}{{\bar r}}
\newcommand{\trc}{{\text{\rm Tr}}}
\newcommand{\tq}{{\bar q}}

\newcommand{\wti}{\widetilde}
\newcommand{\zz}{{\mathbb Z}}

\numberwithin{equation}{section}
\numberwithin{figure}{section}
\numberwithin{table}{section}

\begin{document}

\begin{abstract}
In this paper we consider a family of abstract simplicial complexes
which we call immediate snapshot complexes. Their definition is
motivated by theoretical distributed computing. Specifically, these
complexes appear as protocol complexes in the general immediate
snapshot execution model.  

In order to define and to analyze the immediate snapshot complexes we
use the novel language of witness structures. We develop the
rigorous mathematical theory of witness structures and use it to
prove several combinatorial as well as topological properties of the
immediate snapshot complexes. In particular, we prove that these
complexes are simplicially homeomorphic to simplices.
\end{abstract}

\maketitle

\section{The motivation for the study of immediate snapshot protocol complexes}

One of the core computational models, which is used to understand the
shared-memory communication between a finite number of processes is
the so-called {\it immediate snapshot read/write model}. In this
model, a~number of processes are set to communicate by means of
a~shared memory.  Each process has an assigned register, and each
process can perform two types of operations: {\it write} and {\it
  snapshot read}. The write operation simply writes the entire state
of the process into its assigned register; the snapshot read operation
reads the entire memory in one atomic step. The order in which a
process performs these operations is controlled by the distributed
protocol, whose execution is asynchronous, satisfying an additional
condition. Namely, we assume that at each step a group of processes
gets active. First this group simultaneously writes its values to the
memory, then it simultenously performs a snapshot read. This way, each
execution can be encoded by a sequence of groups of processors which
become active at each turn. More details on this computational model,
the associated protocol complexes and its equivalence with other
models can be found in a recent book~\cite{HKR}.

In this paper we consider the distributed protocols for $n+1$
processes indexed $0,\dots,n$, where the protocol of $k$-th processor
says to run $r_k$ rounds and then to stop. Let the associate protocol
complex be called $P(r_0,\dots,r_n)$. Our first contribution is to
give a rigorous purely combinatorial definition of $P(r_0,\dots,r_n)$.
To do this, we introduce new mathematical objects, which we call {\it
  witness structures} and use them as a language to define and to
analyze this family of simplicial complexes. The special case
$r_0=\dots=r_n=1$ corresponds to the so-called {\it standard chromatic
  subdivision} of a~simplex, see~\cite{subd,view}, the cases where
some $r_i\geq 2$ are new.

The simplicial complexes $P(r_0,\dots,r_n)$ are of utmost importance
in the shared-memory communication. We perform a~thorough analysis of
their combinatorial and topological structure. Our main tool is the
canonical decomposition of $P(r_0,\dots,r_n)$, with strata
corresponding to various groups of processes which take the first turn
of the computation. Our main theorem states that each simplicial complex 
 $P(r_0,\dots,r_n)$ is simplicially homeomorphic to an~$n$-simplex.

\section{The language of witness structures} 

\subsection{Some notations}

We let $\zz_+$ denote the set of nonnegative integers
$\{0,1,2,\dots\}$.  For a natural number $n$ we shall use $[n]$ to
denote the set $\{0,\dots,n\}$.

For a~finite subset $S\in\zz_+$, such that $|S|\geq 2$, we let $\smax
S$ denote the {\it second} largest element $\max(S\setminus\{\max
S\})$.  For a family of finite sets $(S_i)_{i\in I}$,
$S_i\subset\zz_+$, we let $\max(S_i)_{i\in I}$ be the short-hand
notation for $\max_{i\in I}(\max S_i)=\max(\bigcup_{i\in I}S_i)$.

For a set $S$ and an element $a$, we set 
\[\chi(a,S):=
\begin{cases}
1, & \text{ if } a\in S;\\
0, & \text {otherwise.} 
\end{cases}\]

Whenever $(X_i)_{i=1}^t$ is a~family of
topological spaces, we set $X_I:=\cup_{i\in I} X_i$.
Also, when no confusion arises, we identify one-element sets with 
that element, and write, e.g., $p$ instead of $\{p\}$.

We shall use $\hookrightarrow$ to denote simplicial inclusions,
$\rightsquigarrow$ to denote simplicial isomorphisms, and we use
${\underset{\cong}\rightarrow}$, and ${\overset{\cong}\rightarrow}$,
to denote homeomorphisms.





\subsection{Round counters}

Our main objects of study, the immediate snapshot complexes, are
indexed by finite tuples of nonnegative integers. We need to be more
specific about the formalism of this indexing set.

\begin{df}
Given a function $\tr:\zz_+\rightarrow\zz_+\cup\{\bot\}$, we consider
the set
\[\supp\tr:=\{i\in\zz_+\,|\,\tr(i)\neq\bot\}.\] 
This set is called the {\bf support set} of $\tr$.

\nin A {\bf round counter} is a function
$\tr:\zz_+\rightarrow\zz_+\cup\{\bot\}$ with a~finite support set.
\end{df}

Obviously, a~round counter can be thought of as an infinite sequence
$\tr=(\tr(0),\tr(1),\dots)$, where, for all $i\in\zz_+$, either
$\tr(i)$ is a nonnegative integer, or $\tr(i)=\bot$, such that only
finitely many entries of $\tr$ are nonnegative integers. We shall
frequently use a~short-hand notation $\tr=(r_0,\dots,r_n)$ to denote
the round counter given by
\[\tr(i)=\begin{cases}
r_i, & \text{ for } 0\leq i\leq n;\\
\bot,& \text{ for } i>n.
\end{cases}\]

\begin{df}
Given a round counter $\tr$, the number $\sum_{i\in\supp\tr}\tr(i)$ is
called the {\bf cardinality} of $\tr$, and is denoted $|\tr|$. The
sets
\[\act\tr:=\{i\in\supp\tr\,|\,\tr(i)\geq 1\}
\text{ and } \pass\tr:=\{i\in\supp\tr\,|\,\tr(i)=0\}\] 
are called the {\bf active} and the {\bf passive} sets of~$\tr$.
\end{df}

\begin{df}
For an arbitrary pair of disjoint finite sets $A,B\subseteq\zz_+$
we define a~round counter $\chi_{A,B}$ given by
\[\chi_{A,B}(i):=\begin{cases}
1, & \text{ if } i\in A;\\
0, & \text{ if } i\in B.
\end{cases}\]

Furthermore, for an arbitrary round counter $\tr$, we set
$\chi(\tr):=\chi_{\act\tr,\pass\tr}$.
\end{df}

We note that $\supp\tr=\supp(\chi(\tr))$. In the paper we shall also
the short-hand notation $\chi_A:=\chi_{A,\es}$.

We define two operations on the round counters. To start with, assume
$\tr$ is a~\pnt and we have a~subset $A\subseteq\zz_+$. We let
$\tr\sm A$ denote the \pnt defined by
\[(\tr\sm A)(i)=\begin{cases}
\tr(i), & \text{ if } i\notin A;\\
\bot,   & \text{ if } i\in A.
\end{cases}\]
We say that the \pnt $\tr\sm A$ is obtained from $\tr$ by the {\it
  deletion} of~$A$. Note that $\supp(\tr\sm A)=\supp(\tr)\setminus S$,
$\act(\tr\sm A)=\act(\tr)\setminus A$, and $\pass(\tr\sm
A)=\pass(\tr)\setminus A$. Furthermore, we have $\chi(\tr\sm
A)=\chi(\tr)\sm A$. Finally, we note for future reference that for
$A\subseteq C\cup D$ we have
\begin{equation}\label{eq:chi2}
\chi_{C,D}\sm A=\chi_{C\sm A,D\sm A}.
\end{equation} 

For the second operation, assume $\tr$ is a~\pnt and we have a~subset
$S\subseteq\act\tr$. We let $\tr\dar S$ denote the \pnt defined by
\[(\tr\dar S)(i)=\begin{cases}
\tr(i), & \text{ if } i\notin S;\\
\tr(i)-1,   & \text{ if } i\in S.
\end{cases}\]
We say that the \pnt $\tr\dar S$ is obtained from $\tr$ by the {\it
  execution} of~$S$. Note that $\supp(\tr\dar S)=\supp\tr$,
$\act(\tr\dar S)=\{i\in\act\tr\,|\, i\notin S, \textrm{ or }\tr(i)\geq
2 \}$, and $\pass(\tr\dar S)=\pass(\tr)\cup\{i\in S\,|\,\tr(i)=1\}$.
However, in general we have $\chi(\tr)\dar S\neq\chi(\tr\dar S)$.

For an arbitrary round pointer $\tr$ and sets $S\subseteq\act\tr$,
$A\subseteq\supp\tr$ we set
\begin{equation}\label{eq:rsa1}
\tr_{S,A}:=(\tr\dar S)\sm A=(\tr\sm A)\dar(S\sm A).
\end{equation}
In the special case, when $A\cap S=\es$, the identity \eqref{eq:rsa1}
specializes to
\begin{equation}\label{eq:rsa2}
\tr_{S,A}:=(\tr\dar S)\sm A=(\tr\sm A)\dar S.
\end{equation}
When $A=\es$, we shall frequently use the short-hand notation $\tr_S$
instead of $\tr_{S,A}$, in other words, $\tr_S=\tr\dar S$. Again, for
future reference, we note that for $S\subseteq C$, we have
\begin{equation}\label{eq:chi3}
\chi_{C,D}\dar S=\chi_{C\sm S,D\cup S}.
\end{equation}

Assume now we are given a~\pnt $\tr$, and let
  $\varphi:\supp\tr\to[|\supp\tr|-1]$ denote the unique
  order-preserving bijection. The \pnt $c(\tr)$ is defined by
\[c(\tr)(i):=
\begin{cases}
\tr(\varphi^{-1}(i)), &\text{ for } 0\leq i\leq|\supp\tr|-1; \\
\bot,&\text{ for } i\geq|\supp\tr|.
\end{cases}\]
We call $c(\tr)$ the {\it canonical form} of $\tr$. Note that
$\supp c(\tr)=[|\supp\tr|-1]$, $|\act(c(\tr))|=|\act\tr|$,
and $|\pass(c(\tr))|=|\pass(\tr)|$.

Let $\cs_{\zz_+}$ denote the group of bijections $\pi:\zz_+\ra\zz_+$, such that
$\pi(i)\neq\pi(i)$ for only finitely many~$i$. This group acts on
the set of all round counters, namely for $\pi\in\cs_{\zz_+}$, and
a \pnt $\tr$ we set $\pi(\tr)(i):=\tr(\pi(i))$.


\subsection{Witness structures}

\begin{df}\label{df:ws}
A~{\bf witness prestructure} is a~sequence of pairs of finite subsets of $\zz_+$, denoted 
$\sigma=((W_0,G_0),\dots,(W_t,G_t))$, with $t\geq 0$,
satisfying the following conditions:
\begin{enumerate}
\item[(P1)] $W_i,G_i\subseteq W_0$, for all $i=1,\dots,t$;
\item[(P2)] $G_i\cap G_j=\emptyset$, for all $0\leq i<j\leq t$;
\item[(P3)] $G_i\cap W_j=\emptyset$, for all $0\leq i\leq j\leq t$.
\end{enumerate}

\noindent
A witness prestructure is called {\bf stable} if the in addition the following
condition is satisfied:
\begin{enumerate}
\item[(S)] if $t\geq 1$, then $W_t\neq\emptyset$.
\end{enumerate}

\noindent
A {\bf witness structure} is a witness prestructure satisfying the
following strengthening of condition (S):
\begin{enumerate}
\item[(W)] the subsets $W_1,\dots,W_t$ are all nonempty.
\end{enumerate}
\end{df}

\begin{figure}[hbt]
\[
\begin{array}{|c|c|c|c|c|}
\hline
W_0 & W_1 & W_2 & \dots & W_t \\ \hline
G_0 & G_1 & G_2 & \dots & G_t \\ 
\hline
\end{array}\]
\caption{Table presentation of a witness (pre)structure.}
\label{table:ws}
\end{figure}

\noindent
It is often useful to depict a~witness prestructure in form of a~table,
see Figure~\ref{table:ws}. Note, that every witness prestructure with $t=0$ 
is a~witness structure. On the other hand, if $W_0=\emptyset$, then conditions 
(P1) and (S) imply that $t=0$. In this case, only the set $G_0$ carries any information, 
and we call this witness structure {\it empty}. 

\begin{df}
We define the following data associated to an arbitrary witness
prestructure $\sigma=((W_0,G_0),\dots,(W_t,G_t))$:
\begin{itemize}
\item the set $W_0\cup G_0$ is called the {\bf support} of $\sigma$ and is
denoted by $\supp\sigma$;
\item the {\bf ghost set} of $\sigma$ is the set
  $G(\sigma):=G_0\cup\dots\cup G_t$;
\item the {\bf active set} of $\sigma$ is the complement of the ghost set
\[A(\sigma):=\supp(\sigma)\setminus G(\sigma)=W_0\setminus(G_1\cup\dots\cup G_t);\]
\item the {\bf dimension} of $\sigma$ is
  \[\dim\sigma:=|A(\sigma)|-1=|W_0|-|G_1|-\dots-|G_t|-1.\]
\end{itemize}
\end{df}

By definition, the dimension of a~witness prestructure $\sigma$ is
between $-1$ and $|\supp\sigma|-1$. Let us analyze witness structures
of special dimensions. To start with, if $\dim(\sigma)=-1$, then
$A(\sigma)=\emptyset$. In particular, $W_0=\emptyset$, hence
$W_1=\dots,W_t=\emptyset$ and $G_1=\dots=G_t=\emptyset$.  All witness
structures of dimension $-1$ are empty, i.e., of the form
$\sigma=((\emptyset,G_0))$.

Furthermore, it is easy to characterize all witness structures $\sigma$
of dimension~$0$.  In this case, we have $|A(\sigma)|=1$. We let
$\sigma=((W_0,G_0),\dots,(W_t,G_t))$ and let $p$ denote the unique
element of $A(\sigma)$. Then $\sigma$ has dimension~$0$ if and only if
\[W_k\subseteq\{p\}\cup G_{k+1}\cup\dots\cup G_t,
\text{ for all } k=0,\dots,t.\]
 In particular, we must of course have
$W_t=\{p\}$. In such a case, we shall call $p$ the {\it color} of the
strict witness structure~$\sigma$.

At the opposite extreme, a~witness structure 
$\sigma=((W_0,G_0),\dots,(W_t,G_t))$ has dimension $|\supp\sigma|-1$ 
if and only if $G_0=\dots=G_t=\emptyset$. In such a~case, we shall 
frequently use the short-hand notation $\sigma=(W_0,W_1,\dots,W_t)$.

For brevity of some formulas, we set $W_{-1}:=W_0\cup G_0=\supp\sigma$.
Furthermore, we set $R_i:=W_i\cup G_i$, for $i=0,\dots,t$.

\begin{df}\label{df:trc}
For a~prestructure $\sigma$ and an arbitrary $p\in\supp\sigma$, we let
$\trc(p,\sigma)$ denote the set $\{0\leq i\leq t\,|\,p\in R_i\}$, which is
called the {\bf trace} of~$p$. Furthermore, for all $p\in\supp\sigma$, we 
set $\last(p,\sigma):=\max\{i\,|\,p\in W_i\}$, and $M(p,\sigma):=|\trc(p)|$.
\end{df}
When the choice of $\sigma$ is unmbiguous, we shall simply write $\trc(p)$,
$\last(p)$, and $M(p)$. Note furthermore, that if $p\in A(\sigma)$, then 
$\trc(p)=\{0\leq i\leq t\,|\,p\in W_i\}$, while if $p\in G(\sigma)$, then 
$\smax\trc(p)=\{0\leq i\leq t\,|\,p\in W_i\}$ and $p\in G_{\max\trc(p)}$.

To get a~better grasp on the witness structures, as well as operations in them, 
the following alternative approach using traces is often of use.

\begin{df}\label{df:trws}
A~{\bf witness prestructure} is a~pair of finite subsets $A,G\subseteq\zz_+$ 
together with a~family $(\trc(p))_{p\in A\cup G}$ of finite subsets of $\zz_+$, 
satisfying the following two condition:
\begin{itemize}
\item[(T)] $0\in\trc(p)$, for all $p\in A\cup G$.
\end{itemize}

\noindent
A witness prestructure is called {\bf stable} if it satisfies the following
additional condition: 
\begin{itemize}
\item[(TS)] if $A=\emptyset$, then $\trc(p)=\{0\}$, for all $p\in G$,
else 
\[\last(p)_{p\in A}\geq\max(\trc(p))_{p\in G}.\]
\end{itemize}

Set $t:=\last(p)_{p\in A}$. The witness prestructure is called {\bf witness structure}
if the following stregthening of Condition (TS) is satisfied:
\begin{enumerate}
\item[(TW)] {\it for all $1\leq k\leq t$ either there exists $p\in A$ such that
$k\in\trc(p)$, or there exists $p\in G$ such that $k\in\smax\trc(p)$.}
\end{enumerate}
\end{df}

We shall call the form of the presentation of the witness prestructure
described in Definition~\ref{df:trws} its {\it trace form}.

\begin{prop}
The Definitions~\ref{df:ws} and~\ref{df:trws} provide alternative descriptions
of the same mathematical objects.
\end{prop}
\pr The translation between the two descriptions is as follows.
First, assume $\sigma=((W_0,G_0),\dots,(W_t,G_t))$ is a witness prestructure according
to Definition~\ref{df:ws}. Set $A:=A(\sigma)$, $G:=G(\sigma)$, and for each $p\in A\cup G$,
set $\trc(p)$ to be the trace of $p$ as given by Definition~\ref{df:trc}.

Reversely, assume $A$, $G$, and $(\trc(p))_{p\in A\cup G}$. We set
$t:=\max(\trc(p))_{p\in A\cup G}$, and for all $0\leq k\leq t$, we set
\[G_k:=\{p\in G\,|\,k=\max\trc(p)\},\]
\[W_k:=\{p\in A\cup G\,|\,k\in\trc(p)\}\setminus G_k.\]
We leave to the reader to verify that these translations are inverses of each other, 
that they preserve stability, and that they translate witness structures into witness structures.
\qed

\subsection{Operations on witness prestructures}

\subsubsection{Canonical form of a stable witness prestructure} $\,$

\noindent
Any stable witness prestructure can be turned into a witness structure
by means of the following operation.

\begin{df}\label{df:cform}
Assume $\sigma=((W_0,G_0),\dots,(W_t,G_t))$ is an arbitrary stable witness
prestructure. Set $q:=|\{1\leq i\leq t\,|\, W_i\neq\emptyset\}|$.
Pick $0=i_0<i_1<\dots<i_q=t$, such that $\{i_1,\dots,i_q\}=\{1\leq i\leq
t\,|\,W_i\neq\emptyset\}$.  We define the witness structure
$C(\sigma)=((W_0,G_0),(\wti W_1, \wti G_1),\dots,(\wti W_q,\wti G_q))$, which is
called the {\bf canonical form} of~$\sigma$, by setting
\begin{equation}\label{eq:canon}
\wti W_k:=W_{i_k},\quad 
\wti G_k:=G_{i_{k-1}+1}\cup\dots\cup G_{i_k},\text{ for all } k=1,\dots,q,
\end{equation}
\end{df}

\begin{figure}[hbt]
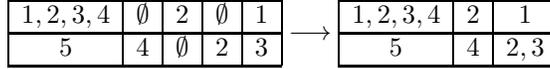

\[
\begin{array}{|c|c|c|c|c|}
\hline
1,2,3,4 & \emptyset & 2 & \emptyset & 1 \\ \hline
5       &  4  &     \emptyset & 2 & 3  \\ 
\hline
\end{array}
\longrightarrow
\begin{array}{|c|c|c|}
\hline
1,2,3,4 & 2 & 1 \\ \hline
5 & 4 & 2,3 \\ 
\hline
\end{array}\]
\caption{A stable witness prestructure and its canonical form.}
\label{table:ws2}
\end{figure}

The construction in Definition~\ref{df:cform} is illustrated 
by Figure~\ref{table:ws2}.

\begin{prop} \label{prop:call}
Assume $\sigma$ is an arbitrary stable witness prestructure.
\begin{itemize}
\item[(a)] The canonical form of $\sigma$ is a~well-defined witness structure.
\item[(b)] We have $C(\sigma)=\sigma$ if and only if $\sigma$
is a~witness structure itself.
\item[(c)] We have  $\supp(C(\sigma))=\supp(\sigma)$,
$A(\sigma)=A(C(\sigma))$, $G(\sigma)=G(C(\sigma))$, and
$\dim(\sigma)=\dim(C(\sigma))$.
\end{itemize}
\end{prop}

\pr Assume $\sigma=((W_0,G_0),\dots,(W_t,G_t))$, $q$ and
$i_1,\dots,i_q$ as in the Definition~\ref{df:cform}, and
$c(\sigma)=((W_0,G_0),(\wti W_1,\wti G_1),\dots,(\wti W_q,\wti G_q))$. 

To prove (a) note first that all the set involved are finite subsets 
of $\zz_+$. Conditions (P1) and (P2) for $C(\sigma)$ follow immediately 
from the correspoding conditions on~$\sigma$. To see (P3), pick some 
$p\in\wti G_k$. Then there exists a~unique $j$, such that 
$i_{k-1}<j\leq i_k$ and $p\in G_j$. Then $p\notin W_j\cup\dots\cup W_t$, 
but $W_j\cup\dots\cup W_t=W_{i_k}\cup\dots\cup W_{i_q}$, hence 
$p\notin\wti W_k\cup\dots\cup\wti W_q$. Finally, to see (W) note that
$W_{i_k}\neq\emptyset$ for all $k=1,\dots,q$, hence $\wti W_k\neq\emptyset$. 

To prove (b) note that if $\sigma$ is a~witness structure, then 
$W_1,\dots,W_t\neq\emptyset$, hence $q=t$, $i_k=k$, for $k=1,\dots,t$. 
It follows that $\wti W_k=W_k$, $\wti G_k=G_k$, for all $k=1,\dots,t$. 
Reversely, assume $C(\sigma)=\sigma$, then $q=t$, hence $i_k=k$, 
for all $k=1,\dots,t$, implying $W_1,\dots,W_t\neq\emptyset$.

To prove (c) note that the first pair of sets in $\sigma$ and in $C(\sigma)$ is 
the same, hence $\supp(C(\sigma))=\supp(\sigma)$. By \eqref{eq:canon} we have 
$\wti W_1\cup\dots\cup\wti W_q=W_1\cup\dots\cup W_t$, and $\wti
G_1\cup\dots\cup\wti G_q=G_1\cup\dots\cup G_t$, hence
$A(\sigma)=A(C(\sigma))$. The other two equalities follow. 
\qed
\vspace{5pt}


\subsubsection{Stabilization of witness prestructures} $\,$

\noindent
Any witness prestructure can be made stable using the following operation.

\begin{df} \label{df:st}
Let $\sigma=((W_0,G_0),\dots,(W_t,G_t))$ be a~witness prestructure, 
and $S\subseteq A(\sigma)$. Set 
\[q:=\max\{0\leq i\leq t\,|\,R_i\not\subseteq S\cup G(\sigma)\}.\]
The {\bf stabilization} of $\sigma$ is the witness prestructure $\st_S(\sigma)$
whose trace form is $A(\sigma)\setminus S$, $G(\sigma)\cup S$, 
$(\trc(p)|_{\{0,\dots,q\}})_{p\in\supp\sigma}$. 
\end{df}
\noindent In the special case $S=\es$ we can simply talk about the 
stabilization of a~witness prestructure.

The following three properties provide an equivalent recursive definition
of stabilization.
\begin{enumerate}
\item[(1)] If $t=0$, then $\st_S(\sigma)=((W_0\setminus S,G_0\cup S))$.
\item[(2)] If $t\geq 1$ and $W_t\subseteq S$, then 
\[\st_S(\sigma)=\st_{S\cup G_t}(((W_0,G_0),\dots,(W_{t-1},G_{t-1}))).\]
\item[(3)] If $t\geq 1$ and $W_t\not\subseteq S$, then the trace form
of $\st_S(\sigma)$ is $A(\sigma)\setminus S$, $G(\sigma)\cup S$, 
$(\trc(p))_{p\in A\cup G}$.
\end{enumerate}

Assume now that $\st_S(\sigma)=((\wti W_0,\wti G_0),\dots,(\wti W_q,\wti G_q))$.
By Definition~\ref{df:st} we have $\wti W_i\cup\wti G_i=R_i$, for all $0\leq i\leq q$.
Hence, for some sets $J_0,\dots,J_q$ we have 
\begin{equation}\label{eq:tfst}
\st_S(\sigma)=((W_0\sm J_0,G_0\cup J_0),\dots,(W_q\sm J_q,G_q\cup J_q)).
\end{equation}
We shall refer to~\eqref{eq:tfst} as the table form of $\st_S(\sigma)$.
The sets $J_i$ are explicitely described by the following formula:
\[J_i:=(W_i\setminus(W_{i+1}\cup\dots\cup W_q))\cap(S\cup G(\sigma)).\]

\begin{figure}[hbt]
\[
\begin{array}{|c|c|c|c|c|c|c|}
\hline
1,2,3,4,5 & 1   & 3,4,5 & 2,3 & 1 & 1 & \es  \\ \hline
\es       & \es & \es   & \es & 3 & 2 & 1 \\ 
\hline
\end{array}
\longrightarrow
\begin{array}{|c|c|c|}
\hline
1,3,4,5 & \es & 4,5 \\ \hline
      2 & 1   &  3\\ 
\hline
\end{array}\]
\caption{Stabilizing a witness prestructure for $S=\{\emptyset\}$.}
\label{table:ws4}
\end{figure}

\begin{prop}\label{prop:stall}
Assume as before that we are given a~strict witness structure $\sigma$, 
and $S\subsetneq A(\sigma)$. The witness prestructure $\st_S(\sigma)$ is 
well-defined and stable. It satisfies the following properties:
\begin{itemize}
\item $\supp(\st_S(\sigma))=\supp\sigma$;
\item $G(\st_S(\sigma))=G(\sigma)\cup S$;
\item $A(\st_S(\sigma))=A(\sigma)\sm S$;
\item $\dim\st_S(\sigma)=\dim\sigma-|S|$.
\end{itemize}
\end{prop}
\pr Straightforward verification.
\qed

\vspace{5pt}

\noindent
The following property if the stabilization will be very useful later on.

\begin{prop} 
\label{prop:stst}
Assume $\sigma$ is a witness prestructure, and $S,T\subseteq A(\sigma)$, 
such that $S\cap T=\es$. Then we have
\begin{equation}\label{eq:stst}
\st_T(\st_S(\sigma))=\st_{S\cup T}(\sigma).
\end{equation}
\end{prop}

\pr Let $\sigma'=\st_T(\st_S(\sigma))$ and $\sigma''=\st_{S\cup T}(\sigma)$.
To show that $\sigma'=\sigma''$ we compare their trace forms. 
To start with, by Definition~\ref{df:st} we have $\supp\sigma'=\supp\sigma$
and $\supp\sigma''=\supp\sigma$. Furthermore, 
$A(\sigma'')=A(\sigma)\setminus(S\cup T)$, and 
$A(\sigma')=A(\st_S(\sigma))\setminus T=(A(\sigma)\setminus S)\setminus T$, 
hence $A(\sigma')=A(\sigma'')$ and $G(\sigma)=G(\sigma'')$.

It remains to show that the traces of the elements from $\supp\sigma$ are 
truncated at the same index in $\sigma'$ and in~$\sigma''$. For $\sigma''$ 
the traces are truncated at 
$q=\max\{0\leq i\leq t\,|\,R_i\not\in S\cup T\cup G(\sigma)\}$. 
On the other hand, to obtain $\st_S(\sigma)$ we truncate at 
$q'=\max\{0\leq i\leq t\,|\,R_i\not\subseteq S\cup G(\sigma)\}$.
Assume $\st_S(\sigma)=((\wti W_0,\wti G_0),\dots,(\wti W_{q'},\wti G_{q'}))$. 
We have $\wti W_i\cup\wti G_i=R_i$, for all $0\leq i\leq q'$. To obtain 
$\sigma'$ from $\st_S(\sigma)$ we now truncate the traces in $\st_S(\sigma)$ 
at $q''=\max\{0\leq j\leq q'\,|\,R_j\not\subseteq T\cup G(\st_S(\sigma))\}$. 
Since $q'\geq q$, and $G(\st_S(\sigma))=G(\sigma)\cup S$,
we obtain $q=q''$. It follows that $\sigma'=\sigma''$.
\qed 

\subsubsection{Ghosting operation on the witness structures} $\,$

\vspace{5pt}

\noindent
We are now ready to define the main operation on witness structures.

\begin{figure}[hbt]
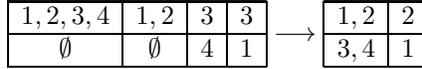

\[
\begin{array}{|c|c|c|c|c|}
\hline
1,2,3,4   & 1,2 & 3 & 3  \\ \hline
\emptyset & \emptyset & 4 & 1 \\ 
\hline
\end{array}
\longrightarrow
\begin{array}{|c|c|}
\hline
1,2 & 2 \\ \hline
3,4 & 1 \\ 
\hline
\end{array}\]
\caption{Ghosting a witness structure for $S=\{3\}$.}
\label{table:ws3}
\end{figure}

\begin{df}\label{df:go}
We define $\Gamma_S(\sigma):=C(\st_S(\sigma))$. We say that
$\Gamma_S(\sigma)$ is obtained from $\sigma$ {\bf by ghosting~$S$}.
\end{df}
\noindent The ghosting operation is illustrated on
Figure~\ref{table:ws3}. When $S=\{p\}$, we shall simply write
$\Gamma_p(\sigma)$.

Clearly, we have $\Gamma_\es(\sigma)=\sigma$. As the next step,
if $S=\{p\}$, i.e., we are ghosting a single element, the situation is
not quite straightforward, though several special cases can be
formulated simpler.

Let $l:=\last(p)$. If $|W_l|\geq 2$, then the situation is much
simpler indeed. In this case $J_i=\emptyset$, for all $i\neq l$, 
while $J_l=\{p\}$. Accordingly, we get 
\begin{multline*}
\Gamma_p(\sigma)=\\ ((W_0,G_0),\dots,(W_{l-1},G_{l-1}),
(W_l\setminus\{p\},G_l\cup\{p\}),(W_{l+1},G_{l+1}),\dots,(W_t,G_t)).
\end{multline*}

The situation is slightly more complex if $|W_l|=1$, i.e.,
$W_l=\{p\}$.  Assume that $l\leq t-1$. Then, we still have 
$J_i=\es$, for all $i\neq l$, and $J_l=\{p\}$. 
The difference now is that 
\[\wti G=((W_0,G_0),\dots,(W_{l-1},G_{l-1}),
(\emptyset,G_l\cup\{p\}),(W_{l+1},G_{l+1}),\dots,(W_t,G_t))\]
is now only a~witness prestructure, so in this case we get
\[
\Gamma_p(\sigma)=((W_0,G_0),\dots,(W_{l-1},G_{l-1}),
(W_{l+1},G_l\cup\{p\}\cup G_{l+1}),\dots,(W_t,G_t)).
\]
Once $l=t$, i.e., $W_t=\{p\}$, we will need the full strength of the
Definition~\ref{df:go}.

The situation is similar if $|S|\geq 2$. For each element $s\in S$ we
set $l(s):=\last(s)$. As long as each $W_{l(S)}$ contains elements
outside of $S$, all that happens is that each element $s\in S$ gets
moved from $W_{l(S)}$ to $G_{l(S)}$. Once this is not true, a more
complex construction is needed.
 
\begin{prop}\label{prop:gall}
Assume as before that we are given a~strict witness structure
$\sigma=((W_0,G_0),\dots,(W_t,G_t))$, and $S\subsetneq A(\sigma)$. The
construction in Definition~\ref{df:go} is well-defined, and yields
a~witness structure $\Gamma_S(\sigma)$, satisfying the following
properties:
\begin{itemize}
\item $\supp(\Gamma_S(\sigma))=\supp\sigma$;
\item $G(\Gamma_S(\sigma))=G(\sigma)\cup S$;
\item $A(\Gamma_S(\sigma))=A(\sigma)\sm S$;
\item $\dim\Gamma_S(\sigma)=\dim\sigma-|S|$.
\end{itemize}
\end{prop}
\pr All equalities follow from the Propositions~\ref{prop:call} and~\ref{prop:stall}. 
\qed

\begin{rem}\label{rem:tr}
For future reference we make the following observation. 
Let $\sigma=((W_0,G_0),\dots,(W_t,G_t))$, and $p,q\in\supp\sigma$.
We always have $M(q,\Gamma_p(\sigma))=M(q,\sigma)$, except for one
single case: namely, when $p=q$ and $W_t=\{p\}$, we have the strict 
inequality $M(p,\Gamma_p(\sigma))<M(p,\sigma)$.
\end{rem}


\begin{lemma} \label{lm:1}
Assume $\sigma$ is a~stable prestructure, and $S\subseteq A(\sigma)$, 
then we have $C(\st_S(\sigma))=C(\st_S(C(\sigma)))$, or expressed functorially 
$C\circ\st_S\circ C=C\circ\st_S$.
\end{lemma}
\pr Assume $\sigma=((W_0,G_0),(W_1,G_1),\dots,(W_t,G_t))$, and set 
$\wti S:=S\cup G(\sigma)$. For appropriately chosen $q$ and 
$0=i_0<i_1<\dots<i_q$ we have
\[C(\sigma)=((W_{i_0},\wti G_{i_0}),(W_{i_1},\wti G_{i_1}),\dots,
(W_{i_q},\wti G_{i_q})),\] 
where $\wti G_{i_k}=\bigcup\limits_{\alpha=i_{k-1}+1}^{i_k} G_\alpha$, 
for $k=0,\dots,q$. 

Set $r:=\max\{1\leq k\leq q\,|\,W_{i_k}\not\subseteq\wti S\}$, and 
$J_k=(W_{i_k}\setminus\bigcup\limits_{\alpha=k+1}^q W_{i_\alpha})\cap\wti S$, 
for $0\leq k\leq r$. Then $i_r:=\max\{1\leq j\leq t\,|\,W_j\not\subseteq\wti S\}$ 
and $J_k=(W_{i_k}\setminus\bigcup\limits_{\alpha=i_k+1}^t W_\alpha)\cap\wti S$, 
for $0\leq k\leq r$, since $W_j=\es$ whenever $j\not\in\{i_0,i_1,\dots,i_q\}$.
It follows that 
\[\st_S(C(\sigma))=((W_{i_0}\setminus J_0,\wti G_{i_0}\cup J_0),
(W_{i_1}\setminus J_1,\wti G_{i_1}\cup J_1),\dots,
(W_{i_r}\setminus J_r,\wti G_{i_r}\cup J_r)).\]
On the other hand, we have $\st_S(\sigma)=((W'_0,G'_0),(W'_1,G'_1),\dots,
(W'_{i_r},G'_{i_r}))$, where
\begin{equation}\label{eq:wgj}
W'_j=\begin{cases}
W_{i_k}\setminus J_k, & \text{ if } j=i_k, \\
\es, & \text{ otherwise};
\end{cases} \qquad
G'_j=\begin{cases}
G_{i_k}\cup J_k, & \text{ if } j=i_k, \\
G_j, & \text{ otherwise}.
\end{cases}
\end{equation}

Set $d:=|\{1\leq k\leq r\,|\,W_{i_k}\setminus J_k\neq\es\}|$, 
then~\eqref{eq:wgj} implies that we also have 
$d=|\{1\leq k\leq i_r\,|\,W'_k\neq\es\}|$. This means that 
$C(\st_S(\sigma))$ and $C(\st_S(C(\sigma)))$ have the same length.

For the appropriate choice of $0=a(0)<a(1)<\dots<a(d)=r$ we have 
\[\{a(1),\dots,a(d)\}=\{1\leq k\leq r\,|\,W_{i_k}\setminus J_k\neq\es\}.\]
Assume $C(\st_S(C(\sigma)))=((V_0,H_0),\dots,(V_d,H_d))$, then we have
$V_k=W_{i_{a(k)}}\setminus J_{a(k)}$,
\begin{equation}\label{eq:hk}
H_k=\bigcup\limits_{\alpha=a(k-1)+1}^{a(k)}(\wti G_{i_\alpha}\cup J_\alpha)=
\bigcup\limits_{\alpha=a(k-1)+1}^{a(k)}\wti G_{i_\alpha}\cup
\bigcup\limits_{\alpha=a(k-1)+1}^{a(k)} J_\alpha,
\end{equation}
for $0\leq k\leq d$. 

Assume now that $C(\st_S(\sigma))=((V'_0,H'_0),\dots,(V'_d,H'_d))$. Note that
\[\{i_{a(1)},\dots,i_{a(d)}\}=\{1\leq k\leq i_r\,|\,W'_k\neq\es\},\] 
hence, for $0\leq k\leq d$, we get
$V'_k=W'_{i_{a(k)}}=W_{i_{a(k)}}\setminus J_{a(k)}$, and 
\[H'_k=\bigcup\limits_{\alpha=i_{a(k-1)}+1}^{i_{a(k)}}G'_\alpha=
\bigcup\limits_{\alpha=i_{a(k-1)}+1}^{i_{a(k)}}G_\alpha\cup
\bigcup\limits_{\alpha=a(k-1)+1}^{a(k)}J_\alpha,\]
where the last equality is a consequence of~\eqref{eq:wgj}. 
Combining the identity
\[\bigcup\limits_{\alpha=a(k-1)+1}^{a(k)}\wti G_{i_\alpha}=
\bigcup\limits_{\alpha=a(k-1)+1}^{a(k)}
\bigcup_{\beta=i_{\alpha-1}+1}^{i_\alpha} G_\beta=
\bigcup\limits_{\beta=i_{a(k-1)}+1}^{i_{a(k)}}G_\beta\]
with~\eqref{eq:hk}, we see that $H_k=H'_k$, for all $0\leq k\leq d$.
\qed

\begin{prop} 
\label{prop:gg}
Assume $\sigma$ is a witness structure, and $S,T\subseteq A(\sigma)$, 
such that $S\cap T=\emptyset$. Then we have
$\Gamma_T(\Gamma_S(\sigma))=\Gamma_{S\cup T}(\sigma)$, expressed 
functorially we have $\Gamma_T\circ\Gamma_S=\Gamma_{S\cup T}$.
\end{prop}

\pr We have 
\[\Gamma_T\circ\Gamma_S=C\circ\st_T\circ C\circ\st_S=
C\circ\st_T\circ\st_S=C\circ\st_{S\cup T}=\Gamma_{S\cup T},\]
where the first and the fourth equalities follow from Definition~\ref{df:go},
the second equality follows from Lemma~\ref{lm:1}, and the third equality 
follows from Proposition~\ref{prop:stst}.
\qed 


\section{Immediate snapshot complexes}

\subsection{Combinatorial definition} $\,$

\vspace{5pt}

\noindent
We now define our main objects of study.

\begin{df}\label{df:ptr}
Assume $\tr$ is a~round counter.
We define a simplicial complex $P(\tr)$ as follows:
\begin{itemize}
\item the simplices of $P(\tr)$ are indexed by witness structures
  $\sigma$ satisfying the following properties:
\begin{enumerate}
\item $\supp\sigma=\supp\tr$;
\item for all $p\in A(\sigma)$, we have $|M(p,\sigma)|=r(p)+1$;
\item for all $p\in G(\sigma)$, we have $|M(p,\sigma)|\leq r(p)+1$.
\end{enumerate}
\item the dimension of the simplex indexed by $\sigma$ is
  $\dim\sigma$, its vertices are
  $\Gamma_{A(\sigma)\setminus\{a\}}(\sigma)$, where $a$ ranges through
  the set $A(\sigma)$.
\end{itemize}
The complex $P(\tr)$ is called the {\bf immediate snapshot complex}
associated to the round counter~$\tr$.
\end{df}

Assume $\tr$ is a round counter, such that $\tr(i)=\bot$ for all
$i\geq n+1$.  In line with our short-hand notation for the round
counters, and in addition skipping a pair of brackets, we shall use an
alternative notation $P(\tr(0),\dots,\tr(n))$ instead of $P(\tr)$.
For every $\sigma\in P(\tr)$ we shall write $V(\sigma)$ to denote the
set of vertices of~$\sigma$. We shall also for brevity often identify
witness structures with the simplices which they are indexing, e.g.,
saying that $\sigma\supseteq\tau$ to indicate that the simplex indexed
by $\sigma$ contains the simplex indexed by~$\tau$.
 
The next proposition checks that the Definition~\ref{df:ptr} yields
a~well-defined simplicial complex, and shows that the ghosting
operation provides the right combinatorial language to describe
boundaries in~$P(\tr)$.

\begin{prop}
\label{prop:b}
Assume $\tr$ is the round counter. 
\begin{itemize}
\item[(1)] The associated immediate snapshot complex $P(\tr)$ is 
a~well-defined simplicial complex.
\item[(2)] Assume $\sigma$ and $\tau$ are simplices of
$P(\tr)$. Then $\tau\subseteq\sigma$ if and only if there exists 
$S\subseteq A(\sigma)$, such that $\tau=\Gamma_S(\sigma)$.
\end{itemize}
\end{prop}

\pr We start by showing (1). We have already observed that the only
witness structure of dimension~$-1$ is the empty one. Since
$\supp\sigma=\supp\tr$, the complex $P(\tr)$ has exactly one simplex
of dimension~$-1$, namely $((\es,\supp\tr))$.

Assume now that the witness structure $\sigma$ indexes a~simplex of
$P(\tr)$. Set $d:=\dim\sigma$, implying that
$A(\sigma)=\{p_0,\dots,p_d\}$ for $p_0<\dots<p_d$, $p_i\in\zz_+$. For
$0\leq i\leq d$, we set
$v_i:=\Gamma_{A(\sigma)\setminus\{p\}}(\sigma)$. We see that the
$d$-dimensional simplex $\sigma$ has $d+1$ vertices, which are all
distinct, since $A(v_i)=p_i$, for $0\leq i\leq d$. Furthermore, it
follows from the Reconstruction Lemma~\ref{lm:rec} that any two
simplices with the same set of vertices are equal, implying that the
simplicial complex $P(\tr)$ is well-defined.

Let us now show (2). To start with, assume $\tau=\Gamma_S(\sigma)$,
for some $S\subseteq A(\sigma)$. By Proposition~\ref{prop:gall} we
have $A(\tau)=A(\sigma)\sm S$.  It follows from
Proposition~\ref{prop:gg} that for every $p\in A(\tau)$ we have
\[\Gamma_{A(\tau)\sm\{p\}}(\tau)=\Gamma_{A(\tau)\sm\{p\}}(\Gamma_S(\sigma))=
\Gamma_{A(\tau)\cup S\sm\{p\}}(\sigma)=\Gamma_{A(\sigma)\sm\{p\}}(\sigma),\] 
hence the set of vertices of $\tau$ is a~subset of the set of vertices of~$\sigma$.

Reversely, assume $V(\tau)\subseteq V(\sigma)$. The same computation as above shows, 
that $V(\Gamma_{\supp\sigma\sm\supp\tau}(\sigma))=\supp\tau$, i.e., $\tau$ and 
$\Gamma_{\supp\sigma\sm\supp\tau}(\sigma)$ have the same set of vertices. 
It follows from the Reconstruction Lemma~\ref{lm:rec} that 
$\tau=\Gamma_{\supp\sigma\sm\supp\tau}(\sigma)$, and so (2) is proved.
\qed

\subsection{The Reconstruction Lemma.} $\,$

\noindent
From the point of view of distributed computing, the vertices of $P(\tr)$
should be thought of as {\it local views} of specific processors. 
In this intuitive picture, the next Reconstruction Lemma~\ref{lm:rec} 
says that any set of local views corresponds to at most one global view.

\begin{lm}\label{lm:rec} {\rm (Reconstruction Lemma).}

\noindent
Assume $\sigma$ and $\tau$ are witness structures, such that the
corresponding $d$-simplices of $P(\tr)$ have the same set of vertices,
then we must have $\sigma=\tau$.
\end{lm}
\pr 
Assume the statement of lemma is not satisfied, and pick a~pair
of $d$-dimensional simplices $\sigma\neq\tau$, such that $V(\sigma)=V(\tau)$,
and $d$ is minimal possible. Obviously, we must have $d\geq 1$.

To start with, the set of vertices defines the support set, so
$\supp\sigma=\supp\tau=\Sigma$. 
Let $p,q\in\Sigma$, then it is easy to check that 
$M(q,\sigma)=M(q,\Gamma_p(\sigma))$, and 
$M(p,\sigma)\leq M(p,\Gamma_p(\sigma))$. This means that the
$\Sigma$-tuples $(M(p,\sigma))_{p\in\Sigma}$ and 
$(M(p,\tau))_{p\in\Sigma}$ are equal.

Let $\sigma=((W_0,G_0),\dots,(W_t,G_t))$. Assume there exists 
$0\leq k\leq t$, such that $|W_k\cap\Sigma|\geq 2$. 
Pick $p,q\in W_k\cap\Sigma$, $p\neq q$, then 
\[\Gamma_p(\sigma)=\begin{array}{|c|c|c|c|c|c|c|}
\hline
W_0 & \dots & W_{k-1} & W_k\sm\{p\}  & W_{k+1} & \dots & W_t \\ \hline
G_0 & \dots & G_{k-1} & G_k\cup\{p\} & G_{k+1} & \dots & G_t \\ 
\hline
\end{array},\]
since $\Gamma_p(\sigma)=\Gamma_p(\tau)$, but $\sigma\neq\tau$, we get
\begin{equation}\label{eq:tau1}
\tau=\begin{array}{|c|c|c|c|c|c|c|c|}
\hline
W_0 & \dots & W_{k-1} & p   & W_k\sm\{p\} & W_{k+1} & \dots & W_t \\ \hline
G_0 & \dots & G_{k-1} & A_p & B_p         & G_{k+1} & \dots & G_t \\ 
\hline
\end{array},\end{equation}
for some $A_p$, $B_p$ such that $A_p\cup B_p=G_k$. Repeating the same argument
with $q$ instead of $p$ we get
\begin{equation}\label{eq:tau2}
\tau=\begin{array}{|c|c|c|c|c|c|c|c|}
\hline
W_0 & \dots & W_{k-1} & q   & W_k\sm\{q\} & W_{k+1} & \dots & W_t \\ \hline
G_0 & \dots & G_{k-1} & A_q & B_q         & G_{k+1} & \dots & G_t \\ 
\hline
\end{array},\end{equation}
for some $A_q$, $B_q$ such that $A_q\cup B_q=G_k$. The
equations~\eqref{eq:tau1} and~\eqref{eq:tau2} contradict each
other. It is thus safe to assume that $|W_k\cap\Sigma|\leq 1$, and
that the same is true for~$\tau$.  An alternative way to phrase the
same condition is to say that $\last(p,\sigma)\neq\last(q,\sigma)$,
and $\last(p,\tau)\neq\last(q,\tau)$, for all $p,q\in\Sigma$.

Set $F:=\{p\in\Sigma\,|\,M(p,\sigma)=M(p,\Gamma_p(\sigma))\}$.  Note
that $F=\{p\in\Sigma\,|\,M(p,\tau)=M(p,\Gamma_p(\tau))\}$.  Using
Remark~\ref{rem:tr}, the previous observation $M(p,\sigma)\leq
M(p,\Gamma_p(\sigma))$ can be strengthened as follows: we know that
$F=\Sigma\sm\{l\}$, for some $l\in\Sigma$. Specifically, $W_t=\{l\}$,
and the last pair of sets in $\tau$ is also $(\{l\},H)$, for some
$H\subseteq G(\tau)$.

Pick $p\in F$ such that $\last(p)=\max_{q\in F}\last(q)$. Assume 
\[\Gamma_p(\sigma)=\begin{array}{|c|c|c|c|c|c|c|}
\hline
W_0 & \dots & W_{k-1} & W_k          & W_{k+1} & \dots & W_t \\ \hline
G_0 & \dots & G_{k-1} & G_k\cup\{p\} & G_{k+1} & \dots & G_t \\ 
\hline
\end{array}.\]
We observe, that $p$ was chosen so that $(W_k\cup\dots\cup W_t)\cap F=\es$. 
We can easily describe the set $\Lambda$ of all $d$-simplices $\gamma$, for which 
$p\in\supp\gamma$ and $\Gamma_p(\gamma)=\Gamma_p(\sigma)$. Set
\[\gamma^p:=\begin{array}{|c|c|c|c|c|c|c|}
\hline
W_0 & \dots & W_{k-1} & W_k\cup\{p\} & W_{k+1} & \dots & W_t \\ \hline
G_0 & \dots & G_{k-1} & G_k          & G_{k+1} & \dots & G_t \\ 
\hline
\end{array},\]
and
\[\gamma_{A,B}:=\begin{array}{|c|c|c|c|c|c|c|c|}
\hline
W_0 & \dots & W_{k-1} & p & W_k & W_{k+1} & \dots & W_t \\ \hline
G_0 & \dots & G_{k-1} & A &  B  & G_{k+1} & \dots & G_t \\ 
\hline
\end{array},\]
where $A\cup B=G_k$. Then $\Lambda=\{\gamma_{A,B}\,|\,A\cup
B=G_k\}\cup\{\gamma^p\}$.  Clearly, $\sigma,\tau\in\Lambda$. We shall
show that $\Gamma_l(\sigma)\neq\Gamma_l(\tau)$.

Assume $A\cup B=G_k$, and pick $\alpha\in W_k$. Then 
\[M(\alpha,\Gamma_l(\gamma^p))=\sum_{i=0}^{k-1}\chi(\alpha,R_i)+1\neq
\sum_{i=0}^{k-1}\chi(\alpha,R_i)= M(\alpha,\Gamma_l(\gamma_{A,B})),\]
hence $\Gamma_l(\gamma^p)\neq\Gamma_l(\gamma_{A,B})$.

Assume now we have further sets $A'$ and $B'$, such that $A'\cup
B'=G_k$, $A\neq A'$.  Without loss of generality, we can assume that
$A\not\subseteq A'$.  Pick now $\alpha\in A\sm A'$. Then
\[M(\alpha,\Gamma_l(\gamma_{A,B}))=\sum_{i=0}^{k-1}\chi(\alpha,R_i)+1\neq
\sum_{i=0}^{k-1}\chi(\alpha,R_i)=M(\alpha,\Gamma_l(\gamma_{A',B'})),\]
hence $\Gamma_l(\gamma_{A,B})\neq\Gamma_l(\gamma_{A',B'})$.

We have thus proved that $\Gamma_l(\sigma)\neq\Gamma_l(\tau)$,
contradicting the choice of $\sigma$ and~$\tau$.  \qed


\section{Some observations on immediate snapshot complexes}

\subsection{Elementary properties and examples} $\,$

\noindent
We start by listing a~few simple but useful properties of the
immediate snapshot complexes~$P(\tr)$.  

First, for an arbitrary point counter $\tr$, we have 
\begin{equation}\label{eq:cfe}
P(\tr)\simeq P(c(\tr)),
\end{equation} 
where $\simeq$ denotes an isomorphism of simplicial complexes.
Specifically, this isomorphism is given by the map
\[\varphi:((W_0,G_0),\dots,(W_t,G_t))\mapsto
((\varphi(W_0),\varphi(G_0)),\dots,(\varphi(W_t),\varphi(G_t))),\]
where $\varphi$ is the unique order-preserving bijection
$\varphi:\supp\tr\rightarrow[|\supp\tr|-1]$. In particular, if \pnts
$\tr$ and $\tq$ have the same canonical form, then the corresponding
immediate snapshot complexes are isomorphic. In other words, the
$\bot$ entries do not matter for the simplicial structure.

In a similar vein, for any \pnt $\tr$, and any permutation
$\pi\in\csn$, the simplicial complex $P(\pi(\tr))$ is isomorphic to
the simplicial complex $P(\tr)$. The isomorphism is given by the map
\[\varphi:((W_0,G_0),\dots,(W_t,G_t))\mapsto
((\pi(W_0),\pi(G_0)),\dots,(\pi(W_t),\pi(G_t))).\]

Let us now look at special round counters. If $\tr=(r)$, then the
simplicial complex $P(\tr)$ is just a~point indexed by the witness
structure $(\underbrace{(0,\es),\dots,(0,\es)}_{r+1})$. Recall, that
the empty simplex of $P(r)$ is indexed by the witness structure
$((\es,0))$.

If $\tr=(\underbrace{0,\dots,0}_{n+1})$, then $P(\tr)$ is isomorphic
to the $n$-simplex $\Delta^{n}$. The simplices of $P(\tr)$ are indexed
by all $((A,B))$ such that $A\cap B=\es$ and $A\cup B=[n]$. The
simplicial isomorphism between $P(\tr)$ and $\da^n$ is given by
$((A,B))\mapsto A$. More generally, if $\tr$ is a round counter such
that $r(i)\in\{\bot,0\}$, for all $i\in\zz_+$, the simplicial complex
$P(\tr)$ is isomorphic with $\Delta^{\supp\tr}$.

Assume now $\tr=(r(0),\dots,r(n))$ and $\tr(n)=0$. Consider related
round counter $\bar q:=(r(0),\dots,r(n-1))$. Consider a cone over
$P(\bar q)$, which we denote $P(\bar q)*\{a\}$, where $a$ is the apex
of the cone. Then we have
\begin{equation}\label{eq:ptr} 
P(\tr)\simeq P(\bar q)*\{a\},
\end{equation} 
with the isomorphism given by 
\[((W_0,G_0),\dots,(W_t,G_t))\mapsto
\begin{cases}
((W_0\setminus\{n\},G_0),\dots,(W_t,G_t))*\{a\},&\textrm{if } n\in W_0;\\
((W_0,G_0\setminus\{n\}),\dots,(W_t,G_t)),&\textrm{if } n\in G_0.
\end{cases}\]

\nin
This observation can be iterated, so that all $0$ entries in $\tr$ are
replaced with the iterated cone construction.

The properties above can be summarized on the intuitive level as
telling us that if we are interested in understanding the simplicial
structure of the complex $P(\tr)$, we may ignore the entries $\bot$
and $0$, and permute the remaining entries as we see fit.

\subsection{The purity of the immediate snapshot complexes} $\,$

\nin
Assume $\sigma=((W_0,G_0),\dots,(W_t,G_t))$ is a~witness structure
which indexes a~simplex of $P(\tr)$. Clearly, we have
$|A(\sigma)|\leq|\supp\tr|$, hence $\dim\sigma\leq|\supp\tr|-1$. It
turns out that every simplex can be extended to the one having 
dimension $|\supp\tr|-1$, implying that immediate snapshot complexes
are always pure.

\begin{prop}\label{prop:pure}
The simplicial complex $P(\tr)$ is pure of dimension $|\supp\tr|-1$.
\end{prop}
\pr Assume $\sigma=((W_0,G_0),\dots,(W_t,G_t))$ is a~witness structure
which indexes a~simplex of $P(\tr)$. For each $p\in G(\sigma)$ we set
$m(p):=r(p)+1-|M(p,\sigma)|$. By construction, we have $m(p)\geq 0$.
Set furthermore $q:=\max_{p\in G(\sigma)} m(p)$,
\[V_i=\{p\in G(\sigma)\,|\,m(p)\geq i\}, \textrm{ for } i=1,\dots,q,\] and
\[\tilde\sigma:=(W_0\cup G_0,W_1\cup G_1,\dots,W_t\cup G_t,V_1,\dots,V_q).\]
We see that $\tilde\sigma$ is a~witness structure: the condition (P1)
says that $V_i\subseteq W_0\cup G_0$, the conditions (P2) and (P3) are
immediate, and condition (W) says that $V_i\neq 0$. Furthermore, we
have $\supp\tilde\sigma=\supp\sigma$, $G(\tilde\sigma)=\es$, and
$A(\tilde\sigma)=\supp\sigma=A(\sigma)\cup G(\sigma)$. For all
$\sigma\in A(\sigma)$ we have
$|M(p,\tilde\sigma)|=|M(p,\sigma)|=r(p)+1$, while for all $\sigma\in
G(\sigma)$ we have $|M(p,\tilde\sigma)|=|M(p,\sigma)|+m(p)=r(p)+1$. We
conclude that $\tilde\sigma$ indexes simplex of $P(\tr)$. Clearly,
$\dim\tilde\sigma=|\supp\sigma|-1$. Finally, we have
$\Gamma(\tilde\sigma,G(\sigma))=\sigma$, so
$\tilde\sigma\subseteq\sigma$ and hence $P(\tr)$ is pure of
dimension~$|\supp\tr|-1$. \qed

\subsection{Immediate snapshot complexes of dimension $1$} $\,$

\nin It follows from the above, that $\dim P(\tr)=0$ if and only if
$|\supp\tr|=1$, meaning that $P(\tr)$ is a~point. Assume now $\dim
P(\tr)=1$. In this case, we have $|\supp\tr|=2$. By \eqref{eq:cfe}, up
to the simplicial isomorphism, we can assume that $\tr=(m,n)$,
$m,n\geq 0$. 

\begin{prop}
For any integers $m,n\geq 0$, the simplicial complex
$P(m,n)$ is a~subdivided interval. 
\end{prop}
\pr The simplicial complex $P(m,n)$ is a pure $1$-dimensional complex.
Hence, it is enough to directly verify that all vertices have valency
$2$, except for the vertices $((0,1),(0,\es),\dots,(0,\es))$ and
$((1,0),(1,\es),\dots,(1,0))$, which have valency~$1$. \qed

\vskip5pt

Let $f(m,n)$ denote the number of $1$-simplices in $P(m,n)$. This number
completely describes the complex $P(m,n)$.

\begin{prop}
The numbers $f(m,n)$ satisfy the recursive relation
\begin{equation}\label{eq:fmn}
f(m,n)=f(m,n-1)+f(m-1,n)+f(m-1,n-1),\quad\forall m,n\geq 1,
\end{equation}
with the boundary conditions $f(m,0)=f(0,m)=1$. The corresponding
generating function
\[F(x,y)=\sum_{m,n=0}^\infty f(m,n)x^m y^n\]
is given by the following explicit formula:
\begin{equation}
F(x,y)=\dfrac{1}{1-x-y-xy}.
\end{equation}	
\end{prop}
\pr Multiply~\eqref{eq:fmn} with $x^ny^n$ and sum over all $m$, $n$.
\qed

\subsection{Number of simplices of maximal dimension in an immediate 
snapshot complex.} $\,$

\nin For arbitrary nonnegative integers $m_0,\dots,m_n$ we let
$f(m_0,\dots,m_n)$ denote the number of top-dimensional simplices in
$P(m_0,\dots,m_n)$. Note that
$f(m_0,\dots,m_n)=f(m_{\pi(0)},\dots,m_{\pi(n)})$ for any
$\pi\in\cs_{[n]}$.
\begin{prop}
We have $f(m_0,\dots,m_{n-1},0)=f(m_0,\dots,m_{n-1})$. Furthermore, if
$m_0,\dots,m_n\geq 1$, we have
\begin{equation}\label{eq:fmn2}
f(m_0,\dots,m_n)=\sum_{\es\neq S\subseteq[n]}f(m_0^S,\dots,m_n^S),
\end{equation}
where 
\[m_k^S=\begin{cases}
m_k-1, & \textrm{ if }k\in S;\\
m_k, & \textrm{ if }k\notin S.
\end{cases}\]
\end{prop}
\pr Immediate consequence of the canonical decomposition
of~$P(m_0,\dots,m_n)$.  \qed

\subsection{Standard chromatic subdivision as immediate snapshot complex} $\,$
\label{ssect:scd}

\nin The {\it standard chromatic subdivision} of an $n$-simplex,
denoted $\chi(\da^n)$, is a~prominent and much studied structure in
distributed computing. We refer to \cite{HKR, HS} for distributed
computing background, and to \cite{subd,view} for the analysis of its
simplicial structure. 

In particular the following combinatorial description of $\chi(\da^n)$
has been given in \cite{view}.  The top dimensional simplices of
$\chi(\Delta^n)$ are indexed by ordered tuples of disjoint sets
$(B_1,\dots,B_t)$ such that $B_1\cup\dots\cup B_t=[n]$. The lower
dimensional simplices of $\chi(\Delta^n)$ are indexed by pairs of
tuples of non-empty sets $((B_1,\dots,B_t)(C_1,\dots,C_t))$, such that
$B_i$'s are disjoint subsets of $[n]$, and $C_i\subseteq B_i$ for
all~$i$. For brevity, set $P_n:=P(\underbrace{1,\dots,1}_{n+1})$.

\begin{prop}
The immediate snapshot complex $P_n$ and
the standard chromatic subdivision of an $n$-simplex are isomorphic as
simplicial complexes. Explicitely, the isomorphism can be given by
\begin{equation}\label{eq:bc}
((B_1,\dots,B_t)(C_1,\dots,C_t))\mapsto
\begin{array}{|c|c|c|c|c|}
\hline
W_0 & C_1        & C_2        & \dots & C_t \\ \hline
[n]\sm W_0 & B_1\sm C_1 & B_2\sm C_2 & \dots & B_t\sm C_t \\ 
\hline
\end{array},
\end{equation}
where $W_0=B_1\cup\dots\cup B_t$.
\end{prop}
\pr 
\qed

Note that \eqref{eq:bc} yields a~direct description of the simplicial structure of
$P_n$, namely the simplices of $P_n$ are indexed by all witness structures 
$\sigma=((W_0,G_0),\dots,(W_t,G_t))$ satisfying the following conditions:
\begin{enumerate} 
	\item [(1)] $W_0\cup G_0=[n]$;
	\item [(2)] $W_0=W_1\cup\dots\cup W_t\cup G_1\cup\dots\cup G_t$;
	\item [(3)] the sets $W_1,\dots,W_t,G_1,\dots,G_t$ are disjoint.
\end{enumerate}


\section{A canonical decomposition of the immediate snapshot complexes}

\subsection{Definition and examples}

\begin{df}
Assume $\tr$ is a~round counter.
\begin{itemize}
\item For every subset $S\subseteq\act\tr$, let $Z_S$ denote the set
  of all simplices $\sigma=((W_0,G_0),\dots,(W_t,G_t))$, such that
  $S\subseteq G_1$.
\item For every pair of subsets $A\subseteq S\subseteq\act\tr$, let
  $Y_{S,A}$ denote the set of all simplices
  $\sigma=((W_0,G_0),\dots,(W_t,G_t))$, such that $R_1=S$ and
  $A\subseteq G_1$. Furthermore, set $X_{S,A}:=Y_{S,A}\cup Z_S$
\end{itemize} 
\end{df}

We shall also use the following short-hand notation:
$X_S:=X_{S,\emptyset}$.  On the other extreme, clearly $Z_S=X_{S,S}$
for all $S$.  When $A\not\subseteq S$, we shall use the convention
$Y_{S,A}=\es$.  Note, that in general the sets $Y_{S,A}$ need not be
closed under taking boundary.

\begin{prop}
The sets $X_{S,A}$ are closed under taking boundary, hence form
simplicial subcomplexes of $P(\tr)$.
\end{prop}
\pr Let $\sigma=((W_0,G_0),\dots,(W_t,G_t))$ be a~simplex in
$X_{S,A}$, and assume $\tau\subset\sigma$. By Proposition~\ref{prop:b}
there exists $T\subseteq A(\sigma)$, such that
$\tau=\Gamma_T(\sigma)$. By Proposition~\ref{prop:gg} it is enough to
consider the case $|T|=1$, so assume $T=\{p\}$, and let $\tau=((\wti
W_0,\wti G_0),\dots,(\wti W_{\tilde t},\wti G_{\tilde t}))$.

By definition of $X_{S,A}$ we have either $\sigma\in Z_S$ or
$\sigma\in Y_{S,A}$.  Consider first the case $\sigma\in Z_S$, so
$S\subseteq G_1$. Since $\wti G_1\supseteq G_1$, we have $\tau\in
Z_S$.

Now, assume $\sigma\in Y_{S,A}$. This means $W_1\cup G_1=S$ and
$A\subseteq G_1$. Again $\wti G_1\supseteq G_1$ implies
$A\subseteq\wti G_1$.  \qed

\vspace{5pt}

\noindent
In particular, $X_S$ and $Z_S$ are simplicial subcomplexes of
$P(\tr)$, for all $S$. When we are dealing with several round
counters, in order to avoid confusion, we shall add $\tr$ to the
notations, and write $X_{S,A}(\tr)$, $X_S(\tr)$, $Y_{S,A}(\tr)$,
$Z_S(\tr)$. We shall also let $\alpha_{S,A}(\tr)$ denote the inclusion
map 
\[\alpha_{S,A}(\tr):X_{S,A}(\tr)\hookrightarrow P(\tr).\]


\subsection{The strata of the canonical decomposition as immediate snapshot complexes}

\begin{prop}\label{prop:strata}
Assume $A\subseteq S\subseteq\act\tr$, then there exists a~simplicial
isomorphism
\[\gamma_{S,A}(\tr):X_{S,A}(\tr)\rightsquigarrow P(\bar r_{S,A}).\] 
\end{prop}
\pr We start by considering the case $A=\es$. Pick an~arbitrary
simplex $\sigma=((W_0,G_0),\allowbreak \dots,(W_t,G_t))$ belonging to $X_S$.  
By the construction of $X_S$, we either have $W_1\cup G_1=S$, or
$S\subseteq G_1$. If $W_1\cup G_1=S$, then set
\[\gamma_S(\sigma):=\begin{array}{|c|c|c|c|}
\hline
W_0\sm G_1  & W_2 & \dots & W_t \\ \hline
G_0\cup G_1 & G_2 & \dots & G_t \\ 
\hline
\end{array},\]
else $S\subseteq G_1$, in which case we set
\[\gamma_S(\sigma):=\begin{array}{|c|c|c|c|c|}
\hline
W_0\sm  S & W_1      & W_2 & \dots & W_t \\ \hline
G_0\cup S & G_1\sm S & G_2 & \dots & G_t \\ 
\hline
\end{array}.\]

Reversely, assume $\tau=((V_0,H_0),\dots,(V_t,H_t))$ is a~simplex of
$P(\bar r_S)$. Note, that in any case, we have $S\subseteq V_0\cup
H_0$. If $V_0\cap S\neq\es$, we set
\[\rho_S(\tau):=\begin{array}{|c|c|c|c|c|}
\hline
V_0\cup(H_0\cap S) & V_0\cap S & V_1 & \dots & V_t \\ \hline
H_0\sm(H_0\cap S)  & H_0\cap S & H_1 & \dots & H_t \\ 
\hline
\end{array},\]
else $S\subseteq H_0$, and we set
\[\rho_S(\tau):=\begin{array}{|c|c|c|c|c|}
\hline
V_0\cup S & V_1       & V_2 & \dots & V_t \\ \hline
H_0\sm  S & H_1\cup S & H_2 & \dots & H_t \\ 
\hline
\end{array}.\]

It is immediate that $\gamma_S$ and $\rho_S$ preserve supports,
$A(-)$, $G(-)$, and hence also the dimension. Furthermore, we can see
what happens with the cardinalities of the traces. For all elements
$p$ which do not belong to~$S$, the cardinalities of their traces are
preserved. For all elements in $S$, the map $\gamma_S$ decreases the
cardinality of the trace, whereas, the map $\rho_S$ increases it. It
follows that $\gamma_S$ and $\rho_S$ are well-defined as
dimension-preserving maps between sets of simplices.

To see that $\gamma_S$ preserves boundaries, pick a~top-dimensional
simplex $\sigma=(W_0,S,\allowbreak W_1,\dots,W_t)$ in $X_S$ and ghost the
set~$T$. Assume first $S\not\subseteq T$. In this case not all
elements in $S$ are ghosted. Assume now that $S\subseteq T$. This
implies that $\gamma_S$ is well-defined as a~simplicial map.  Finally,
a~direct verification shows that the maps $\gamma_S$ and $\rho_S$ are
inverses of each other, hence they are simplicial isomorphisms.

Let us now consider the case when $A$ is arbitrary. The simplicial
complex $X_{S,A}$ is a~subcomplex of $X_S$ consisting of all simplices
$\sigma$ satisfying the additional condition $A\subseteq G_1$. The
image $\gamma_S(X_{S,A})$ consists of all
$\tau=((V_0,H_0),\dots,(V_t,H_t))$ in $P(\bar r_{S,A})$ satisfying
$A\subseteq H_0$. The map $\Xi:\gamma_S(X_{S,A})\rightarrow P(\bar
r_{S,A})$, taking $\tau$ to $((V_0,H_0\setminus
A),(V_1,H_1),\dots,(V_t,H_t))$, is obviously a~simplicial isomorphism,
hence the composition
$\gamma_{S,A}=\Xi\circ\gamma_S:X_{S,A}\rightarrow P(\bar r_{S,A}))$ is
a~simplicial isomorphism as well.  \qed

\vskip5pt

\nin Note that, in particular,
\[\gamma_{A,A}(\sigma)=\begin{array}{|c|c|c|c|c|}
\hline
W_0\sm A & W_1      & W_2 & \dots & W_t \\ \hline
G_0      & G_1\sm A & G_2 & \dots & G_t \\ 
\hline
\end{array}.\]

\begin{prop}\label{prop:cxs}
Assume $\tr$ is an arbitrary round counter, and $S,A\subset\act\tr$,
such that $S\cap A=\es$, then the following diagram commutes
\begin{equation}\label{cd:xs}
\begin{tikzcd}[column sep=large]
X_{A,A}(\tr)\arrow[hookleftarrow]{r}{i}
\arrow[squiggly]{d}[swap]{\gamma_{A,A}(\tr)}
&X_{S\cup A,A}(\tr)\arrow[squiggly]{rd}{\gamma_{S\cup A,A}(\tr)} \\
P(\tr\sm A)\arrow[hookleftarrow]{r}{\alpha_S(\tr\sm A)}
& X_S(\tr\sm A)\arrow[squiggly]{r}{\gamma_S(\tr\sm A)}
& P(\tr_{S,A}),
\end{tikzcd}
\end{equation}
where $i$ denotes the strata inclusion map.
\end{prop}
\pr To start with, note that $\tr_{S,A}=(\tr\dar S)\sm
A=(\tr\dar(S\cup A))\sm A$, so the diagram \eqref{cd:xs} is
well-defined. To see that it is commutative, pick an arbitrary
$\sigma=((W_0,G_0),\dots,(W_t,G_t))$. We know that either $A\subseteq
G_1$ and $W_1\cup G_1=S\cup A$, or $A\cup S\subseteq G_1$. On one
hand, we have
\[(\gamma_{A,A}(\tr)\circ i)(\sigma)=\begin{array}{|c|c|c|c|c|}
\hline
W_0\sm A & W_1      & W_2 & \dots & W_t \\ \hline
G_0      & G_1\sm A & G_2 & \dots & G_t \\ 
\hline
\end{array}.\]
On the other hand, we have
\[\gamma_{S\cup A,A}(\tr)(\sigma)=
\begin{cases}
\begin{array}{|c|c|c|c|}
\hline
W_0\sm G_1       &  W_2 & \dots & W_t \\ \hline
G_0\cup G_1\sm A &  G_2 & \dots & G_t \\ 
\hline
\end{array}, 
\textrm{ if } A\subseteq G_1, W_1\cup G_1=S\cup A;\\[0.6cm]
\begin{array}{|c|c|c|c|c|}
\hline
W_0\sm(S\cup A) & W_1             & W_2 & \dots & W_t \\ \hline
G_0\cup S       & G_1\sm(S\cup A) & G_2 & \dots & G_t \\ 
\hline
\end{array},
\textrm{ if } A\cup S \subseteq G_1.
\end{cases}\]
Applying $\gamma_S(\tr\sm A)^{-1}$ we can verify that
$\gamma_{A,A}(\tr)\circ i=\alpha_S(\tr\sm A)\circ\gamma_S(\tr\sm
A)^{-1}\circ\gamma_{S\cup A,A}(\tr)$.  \qed

\begin{crl}\label{crl:xaa}
 For any $A\subseteq\act\tr$, we have
\begin{equation}\label{eq:xaa}
X_{A,A}(\tr)=\bigcup_{\es\neq S\subseteq\act\tr\sm A}X_{S\cup A,A}(\tr)=
\bigcup_{A\subset S\subseteq\act\tr}X_{S,A}(\tr).
\end{equation}
\end{crl}
\pr Since $P(\tr\sm A)=\bigcup_{\es\neq S\subseteq\act\tr\sm
  A}X_S(\tr\sm A)$, the equation~\eqref{eq:xaa} is an immediate
consequence of the commutativity of the diagram~\eqref{cd:xs}. \qed

\subsection{The incidence structure of the canonical decomposition} $\,$

\nin
Clearly, $P(\tr)=\cup_S X_S$.  We describe here the complete
combinatorics of intersecting these pieces.

\begin{prop} \label{pr:inc1}
For all pairs of subsets $A\subseteq S\subseteq\supp\tr$ and
$B\subseteq T\subseteq\supp\tr$ we have: $X_{S,A}\subseteq X_{T,B}$ if
and only if at least one of the following two conditions is satisfied:
\begin{itemize}
\item $S=T$ and $B\subseteq A$,
\item $T\subseteq A$.
\end{itemize} 
\end{prop}

\noindent
We remark that it can actually happen that both conditions in
Proposition~\ref{pr:inc1} are satisfied. This happens exactly when
$S=T=A$.

\vspace{5pt}

\noindent
{\bf Proof of Proposition~\ref{pr:inc1}.}  First we show that
$T\subseteq A$ implies $X_{S,A}\subseteq X_{T,B}$.  Take $\sigma\in
X_{S,A}$. If $\sigma\in Z_S$, then we have the following chain of
implications: $S\subseteq G_1\Rightarrow A\subseteq G_1\Rightarrow
T\subseteq G_1\Rightarrow\sigma\in Z_T$. If, on the other hand,
$\sigma\in Y_{S,A}$, we also have $A\subseteq G_1$, implying
$T\subseteq G_1$, hence $\sigma\in Z_T$.

Next we show that if $S=T$ and $B\subseteq A$, then $X_{S,A}\subseteq
X_{S,B}$.  Clearly, we just need to show that $Y_{S,A}\subseteq
X_{S,B}$.  Take $\sigma\in Y_{S,A}$, then we have the following chain
of implications:
\[\begin{cases}R_1=S\\A\subseteq G_1\end{cases}\Rightarrow
\begin{cases}R_1=T\\B\subseteq G_1\end{cases}
\Rightarrow\sigma\in Y_{T,B}.\]
This proves the {\it if} part of the proposition.

To prove the {\it only if} part, assume $X_{S,A}\subseteq X_{T,B}$. If
$S\neq A$, set
\[\tau:=\begin{array}{|c|c|c|c|c|}
\hline
\supp\tr & S\sm A & p_1 & \dots & p_t \\ \hline
\es      & A      & \es & \dots & \es \\ 
\hline
\end{array},\]
else $S=T$, and we set
\[\tau:=\begin{array}{|c|c|c|c|c|}
\hline
\supp\tr & p_1 & p_2 & \dots & p_t \\ \hline
\es      & S   & \es & \dots & \es \\ 
\hline
\end{array},\]
where in both cases $p_1,\dots,p_t$ is a sequence of elements from
$\supp\tr\sm A$, with each element $p$ occurring $\tr(p)$
times. Clearly, in the first case, $\tau\in Y_{S,A}$, and in the
second case $\tau\in Z_S$, hence $\tau\in X_{T,B}=Z_T\cup Y_{T,B}$.
This means that either $T\subseteq A$, or $S=T$ and $B\subseteq A$.
\qed

\begin{lm}\label{lm:yzint}
Assume $A\subseteq S\subseteq\supp\tr$ and
$B\subseteq T\subseteq\supp\tr$. We have
\begin{enumerate}
	\item [(1)] $Z_S\cap Z_T=Z_{S\cup T}$,
	\item [(2)] $Y_{S,A}\cap Z_T=Y_{S,A\cup T}$,
	\item [(3)] $Y_{S,A}\cap Y_{T,B}=\begin{cases}
	Y_{S,A\cup B}, & \textrm{ if } S=T,\\ 
	\es, & \textrm{ otherwise}.\end{cases}$
\end{enumerate}
\end{lm}
\pr To show (1), pick $\sigma\in Z_S\cap Z_T$. We have $S\subseteq
G_1$ and $T\subseteq G_1$, hence $S\cup T\subseteq G_1$, and so
$\sigma\in Z_{S\cup T}$.

To show (2), pick $\sigma\in Y_{S,A}\cap Z_T$. We have $R_1=S$,
$A\subseteq G_1$, and $T\subseteq G_1$. It follows that $R_1=S$ and
$A\cup T\subseteq G_1$, so $\sigma\in Y_{S,A\cup T}$.

Finally, to show (3), pick $\sigma\in Y_{S,A}\cap Y_{T,B}$. On one
hand, $\sigma\in Y_{S,A}$ means $R_1=S$ and $A\subseteq G_1$, on the
other hand, $\sigma\in Y_{T,B}$ means $R_1=T$ and $B\subseteq G_1$. We
conclude that $Y_{S,A}\cap Y_{T,B}=\es$ if $S\neq T$. Otherwise, we
have $R_1=S=T$ and $A\cup B\subseteq G_1$, so $\sigma\in Y_{S,A\cup
  B}$.  \qed

\begin{prop} \label{pr:inc2}
For all pairs of subsets $A\subseteq S\subseteq\supp\tr$ and
$B\subseteq T\subseteq\supp\tr$ we have the following formulae for the
intersection:
\begin{numcases}
{X_{S,A}\cap X_{T,B}=}
X_{S,A\cup B}, & if  $S=T$; \label{inc:1} \\
X_{T,S\cup B}, & if  $S\subset T$; \label{inc:2}\\
Z_{S\cup T}=X_{S\cup T,S\cup T}, & if  
$S\not\subseteq T$ and $T\not\subseteq S$. \label{inc:3}
\end{numcases}
\end{prop}
\pr In general, we have
\begin{multline} \label{eq:xx}
X_{S,A}\cap X_{T,B}=(Z_S\cap Z_T)\cup(Z_S\cap Y_{T,B})\cup(Y_{S,A}\cap Z_T)\cup
(Y_{S,A}\cap Y_{T,B})\\
=\begin{cases}
Z_{S\cup T}\cup Y_{T,S\cup B}\cup Y_{S,T\cup A}\cup Y_{S,A\cup B}, & \text{ if } S=T; \\
Z_{S\cup T}\cup Y_{T,S\cup B}\cup Y_{S,T\cup A}, & \text{ otherwise}.
\end{cases}
\end{multline}

Assume first that $S=T$. In this case $Y_{T,S\cup B}=Y_{S,T\cup
  A}=Z_S$, hence the equation \eqref{eq:xx} translates to $X_{S,A}\cap
X_{T,B}=Z_S\cup Y_{S,A\cup B}=X_{S,A\cup B}$.

Let us now consider the case $S\subset T$. We have $Y_{S,T\cup
  A}=\es$, hence \eqref{eq:xx} translates to $X_{S,A}\cap
X_{T,B}=Z_T\cup Y_{T,S\cup B}=X_{T,S\cup B}$.

Finally, assume $S\not\subseteq T$ and $T\not\subseteq S$. Then
$Y_{T,S\cup B}=Y_{S,T\cup A}=\es$, hence \eqref{eq:xx} says
$X_{S,A}\cap X_{T,B}=Z_{S\cup T}$.  \qed

For convenience we record the following special cases of
Proposition~\ref{pr:inc2}.

\begin{crl}\label{crl:6.6}
For $S\neq T$ we have
\[X_S\cap X_T=\begin{cases}
X_{T,S}, & \text{ if } S\subset T, \\
Z_{S\cup T}, & \text{ otherwise,}
\end{cases}\]
\begin{equation}
\label{eq:xz}
X_S\cap Z_T=Z_{S\cup T}.
\end{equation}
\end{crl}
\pr The first formula is a simple substitution of $A=B=\es$ in
\eqref{inc:2} and \eqref{inc:3}. To see \eqref{eq:xz}, substitute
$A=\es$, $B=T$ in \eqref{inc:2} to obtain
\[X_{S,\es}\cap X_{T,T}=\begin{cases}
X_{T,S\cup T}, & \text{ if } S\subset T \\
Z_{S\cup T}, & \text{ otherwise}
\end{cases}=\begin{cases}
Z_T, & \text{ if } S\subset T \\
Z_{S\cup T}, & \text{ otherwise}
\end{cases}=Z_{S\cup T}.\mqed
\]

\begin{rem}
Corollary~\ref{crl:6.6} implies that every stratum $X_{S,A}$ can be
represented as an intersection of two strata of the type $X_S$, with
only exception provided by the strata $X_{S,S}$, when $|S|=1$.
\end{rem}

\begin{crl}\label{crl:inter}
Assume $S_1,\dots,S_t\subseteq[n]$, such that $S_1\not\subset S_i$,
for all $i=2,\dots,t$. The following two cases describe the
intersection $X_{S_1}\cap\dots\cap X_{S_t}$:
\begin{enumerate}
\item[(1)] if $S_1\supset S_i$, for all $i=2,\dots,t$, then
  $X_{S_1}\cap\dots\cap X_{S_t}=X_{S_1,S_2\cup\dots\cup S_t}$;
\item[(2)] if there exists $2\leq i\leq t$, such that $S_1\not\supset S_i$,
  then $X_{S_1}\cap\dots\cap X_{S_t}=Z_{S_1\cup S_2\cup\dots\cup S_t}
  =X_{S_1\cup S_2\cup\dots\cup S_t,S_1\cup S_2\cup\dots\cup S_t}$.
\end{enumerate}
\end{crl}
\pr Assume first that $S_1\supset S_i$, for all $i=2,\dots,t$. By iterating
\eqref{inc:2} we get
\begin{multline*} X_{S_1}\cap\dots\cap X_{S_t}=X_{S_1,\es}\cap X_{S_2,\es}\cap\dots 
\cap X_{S_t,\es}= X_{S_1,S_2}\cap X_{S_3,\es}\dots\cap
X_{S_t,\es}\\ =X_{S_1,S_2\cup S_3}\cap X_{S_4,\es}\dots\cap
X_{S_t,\es}=\dots=X_{S_1,S_2\cup\dots\cup S_t}.
\end{multline*}
This proves (1). 

To show (2), we can assume without loss of generality, that 
$S_2\not\subset S_1$.  By \eqref{inc:3} we have $X_{S_1}\cap X_{S_2}=Z_{S_1\cup S_2}$.  
By iterating \eqref{eq:xz} we get
\[
Z_{S_1\cup S_2}\cap X_{S_3}\cap\dots\cap X_{S_t}=Z_{S_1\cup S_2\cup
  S_3}\cap X_{S_4}\cap\dots\cap X_{S_t} =X_{S_1\cup S_2\cup\dots\cup
  S_t},
\]
which finishes the proof.
\qed

\subsection{The boundary of the immediate snapshot complexes and its canonical 
decomposition }

\begin{df}\label{df:bv}
Let $\tr$ be an arbitrary round counter, and assume
$V\subset\supp\tr$. We define $B_V(\tr)$ to be the simplicial
subcomplex of $P(\tr)$ consisting of all simplices
$\sigma=((W_0,G_0),\dots,\allowbreak (W_t,G_t))$, satisfying
$V\subseteq G_0$.
\end{df}

The fact that $B_V(\tr)$ is a~well-defined subcomplex of $P(\tr)$ is
immediate from the definition of the ghosting operation. We shall
let $\beta_V(\tr)$ denote the inclusion map
\[\beta_V(\tr):B_V(\tr)\hookrightarrow P(\tr).\]

\begin{prop}
For an arbitrary round counter $\tr$, and any $V\subset\supp\tr$, the map
$\delta_V(\tr)$ given by
\[\delta_V(\tr):((W_0,G_0),\dots,(W_t,G_t))\mapsto((W_0,G_0\sm V),\dots,(W_t,G_t))\]
is a~simplicial isomorphism between simplicial complexes $B_V(\tr)$
and $P(\tr\sm V)$.
\end{prop}
\pr The map $\delta_v(\tr)$ is simplicial, and it has a~simplicial
inverse which adds $V$ to $G_0$.  \qed

\vskip5pt

Given an arbitrary round counter $\tr$, $A\subseteq
S\subseteq\act\tr$, and $V\subset\supp\tr$, such that $S\cap V=\es$,
we set \[X_{S,A,V}(\tr):=X_{S,A}(\tr)\cap B_V(\tr).\] We can use the
notational convention $B_\es(\tr)=P(\tr)$, which is consistent with
Definition~\ref{df:bv}. In this case we get
$X_{S,A,\es}(\tr)=X_{S,A}(\tr)$, fitting well with the previous
notations.

\begin{prop}\label{prop:6.13}
Assume $\tr$ is an arbitrary round counter, $V\subset\supp\tr$,
$A\subseteq S\subseteq\act\tr$, and $V\cap S=\es$. Then there exist
simplicial isomorphisms $\varphi$ and $\psi$ making the following
diagram commute:
\begin{equation}\label{eq:bar}
\begin{tikzcd}
P(\tr)\arrow[hookleftarrow]{r}{\alpha}\arrow[hookleftarrow]{d}{\beta} 
&X_{S,A}(\tr)\arrow[squiggly]{r}{\gamma}\arrow[hookleftarrow]{d}{j} 
&P(\tr_{S,A})\arrow[hookleftarrow]{d}{\beta} \\
B_V(\tr)\arrow[hookleftarrow]{r}{i}
\arrow[squiggly]{d}{\delta} 
&X_{S,A,V}(\tr)\arrow[squiggly]{r}{\varphi}\arrow[squiggly]{d}{\psi} 
&B_V(\tr_{S,A})\arrow[squiggly]{d}{\delta} \\
P(\tr\sm V)\arrow[hookleftarrow]{r}{\alpha} 
&X_{S,A}(\tr\sm V)\arrow[squiggly]{r}{\gamma} 
&P(\bar r_{S\cup V,A\cup V}),
\end{tikzcd}
\end{equation}
where $i$ and $j$ denote inclusion maps.
\end{prop}
\pr Note that $X_{S,A,V}(\tr)$ consists of all simplices
$\sigma=((W_0,G_0),\dots,(W_t,G_t))$, such that $V\subseteq G_0$,
$A\subseteq G_1$, and either $W_1\cup G_1=S$, or $S\subseteq G_1$. The
fact that $V$ and $S$ are disjoint ensures that these conditions do
not contradict each other.  We let $\varphi$ be the restriction of
$\gamma_{S,A}(\tr):X_{S,A}(\tr)\ra P(\tr_{S,A})$ to $X_{S,A,V}(\tr)$.
Furthermore, we let $\psi$ be the restriction of
$\delta_V(\tr):B_V(\tr)\ra P(\tr\sm V)$ to $X_{S,A,V}(\tr)$.  \qed

\vskip5pt

\nin The diagram \eqref{eq:bar} means that we can naturally think
about $X_{S,A,V}(\tr)$ both as $X_{S,A}(\tr\sm V)$ as well as
$B_V(\tr_{S,A})$, or abusing notations we write $B_V\cap
X_{S,A}=X_{S,A}(B_V)=B_V(X_{S,A})$.

\begin{prop}\label{prop:5}
Assume $B\subseteq A\subseteq S\subseteq\act\tr$, then the following
diagram commutes
\begin{equation}\label{cd:b1}
\begin{tikzcd}[column sep=large]
X_{S,B}(\tr)\arrow[hookleftarrow]{r}{i}\arrow[squiggly]{d}[swap]{\gamma_{S,B}(\tr)}
& X_{S,A}(\tr)\arrow[squiggly]{rd}{\gamma_{S,A}(\tr)} \\
P(\tr_{S,B})\arrow[hookleftarrow]{r}{\beta_{A\sm B}(\tr_{S,B})}
& B_{A\sm B}(\tr_{S,B})\arrow[squiggly]{r}{\delta_{A\sm B}(\tr_{S,B})}
& P(\tr_{S,A})
\end{tikzcd}
\end{equation}
where $i$ denotes the inclusion map.
\end{prop}
\pr Take $\sigma=((W_0,G_0),\dots,(W_t,G_t))\in X_{S,A}(\tr)$. 
On one hand we have
\[(\gamma_{S,B}(\tr)\circ i)(\sigma)=
\begin{cases}
\begin{array}{|c|c|c|c|}
\hline
W_0\sm G_1       &  W_2 & \dots & W_t \\ \hline
G_0\cup G_1\sm B &  G_2 & \dots & G_t \\ 
\hline
\end{array}, 
\textrm{ if } W_1\cup G_1=S,\,\,A\subseteq G_1;\\[0.6cm]
\begin{array}{|c|c|c|c|c|}
\hline
W_0\sm S  & W_1      & W_2 & \dots & W_t \\ \hline
G_0\cup S\sm B & G_1\sm S & G_2 & \dots & G_t \\ 
\hline
\end{array},
\textrm{ if } S \subseteq G_1.
\end{cases}
\]
On the other hand, we have 
\[(\gamma_{S,A}(\tr))(\sigma)=
\begin{cases}
\begin{array}{|c|c|c|c|}
\hline
W_0\sm G_1       &  W_2 & \dots & W_t \\ \hline
G_0\cup G_1\sm A &  G_2 & \dots & G_t \\ 
\hline
\end{array}, 
\textrm{ if } W_1\cup G_1=S,\,\,A\subseteq G_1;\\[0.6cm]
\begin{array}{|c|c|c|c|c|}
\hline
W_0\sm S  & W_1      & W_2 & \dots & W_t \\ \hline
G_0\cup S\sm A & G_1\sm S & G_2 & \dots & G_t \\ 
\hline
\end{array},
\textrm{ if } S \subseteq G_1.
\end{cases}
\]
Since applying $\delta_{A\sm B}(\tr_{S,B})^{-1}$ will add $A\sm B$ to
  $G_0\cup G_1\sm A$, resp.\ $G_0\cup S\sm A$, above and $A\subseteq
  S$, $A\subseteq G_1$, we conclude that
\[(\gamma_{S,B}(\tr)\circ i)(\sigma)=(\beta_{A\sm B}(\tr_{S,B})\circ
\delta_{A\sm B}(\tr_{S,B})^{-1}\circ\gamma_{S,A}(\tr))(\sigma).\]
Which is the same as to say that the diagram~\eqref{cd:b1} commmutes.  
\qed

\subsection{The combinatorial structure of the complexes $P(\chi_{A,B})$} $\,$

\nin Let us analyze the simplicial structure of $P(\chi_{A,B})$. Set
$k:=|A|-1$ and $m:=|B|$. By~\eqref{eq:ptr} the simplicial complex
$P(\chi_{A,B})$ is isomorphic to the $m$-fold suspension of
$P(\chi_A)$. On the other hand, we saw in
subsection~\ref{ssect:scd} that $P(\chi_A)$ is isomorphic to the
standard chromatic subdivision of $\Delta^k$. The simplices of the
$m$-fold suspension of $\chi(\da^k)$ (which is of course homeomorphic
to $\da^{m+k}$) are indexed by tuples
$(S,(B_1,\dots,B_t)(C_1,\dots,C_t))$, where $S$ is any subset of $B$,
and the sets $B_1,\dots,B_t,C_1,\dots,C_t$ satisfy the same conditions
as in the combinatorial description of the simplicial structure of
$\chi(\da^k)$.  In line with ~\eqref{eq:bc}, the simplicial
isomorphism between $P(\chi_{A,B})$ and the $m$-fold suspension of
$\chi(\da^k)$ can be explicitely given by
\[(S,(B_1,\dots,B_t)(C_1,\dots,C_t))\mapsto
\begin{array}{|c|c|c|c|}
\hline
W_0            & C_1        & \dots & C_t \\ \hline
(A\cup B)\sm W_0 & B_1\sm C_1 & \dots & B_t\sm C_t \\ 
\hline
\end{array},\]
where $W_0=S\cup B_1\cup\dots\cup B_t$. 
In particular, up to the simplicial
isomorphism, the complex $P(\chi_{A,B})$ depends only on $m$ and~$k$.

In analogy with subsection~\ref{ssect:scd}
the simplices of $P(\chi_{A,B})$ are indexed by all witness structures 
$\sigma=((W_0,G_0),\dots,(W_t,G_t))$ satisfying the following conditions:
\begin{enumerate} 
	\item [(1)] $W_0\cup G_0=A\cup B$;
	\item [(2)] $W_0\cap A=W_1\cup\dots\cup W_t\cup G_1\cup\dots\cup G_t$;
	\item [(3)] the sets $W_1,\dots,W_t,G_1,\dots,G_t$ are disjoint.
\end{enumerate}

It was shown in~\cite{subd} that there is a~homeomorphism
\[\tau_A:P(\chi_A)\,{\underset{\cong}\longrightarrow}\,\Delta^A,\]
such that for any $C\subseteq A$ the following diagram commutes
\begin{equation}\label{cd:tau3}
\begin{tikzcd}[column sep=1.5cm]
P(\chi_A)\arrow[hookleftarrow]{r}{\beta_{A\sm C}(\chi_A)}
\arrow{d}{\cong}[swap]{\tau_A} 
&B_{A\sm C}(\chi_A)\arrow[squiggly]{r}{\delta_{A\sm C}(\chi_A)} 
&P(\chi_C)\arrow{ld}{\cong}[swap]{\tau_C}\\
\da^A\arrow[hookleftarrow]{r}{i}&\da^C
\end{tikzcd}
\end{equation}
where $i:\da^C\hookrightarrow\da^A$ is the standard inclusion map.
In general, given a pair if sets $(A,B)$, we take the $|B|$-fold
suspension of the map $\tau_A$ to produce a~homeomorphism  
\[\tau_{A,B}:P(\chi_{A,B})\underset{\cong}\longrightarrow\Delta^{A\cup B}.\]

\begin{df}\label{df:tau}
When $A\cup B=C\cup D$, we set
\[\tau(\chi_{A,B},\chi_{C,D}):=\tau_{C,D}^{-1}\circ\tau_{A,B},\] 
clearly, we get
a homeomorphism
$\tau(\chi_{A,B},\chi_{C,D}):P(\chi_{A,B})\underset{\cong}\longrightarrow P(\chi_{C,D})$.
\end{df}
We know that this map is a simplicial isomorphism when restricted to
$B_S(\chi_{A,B})$, for all $S\subseteq (A\cap C)\cup (B\cap D)$, i.e.,
we have the following commutative diagram
\begin{equation}
\begin{tikzcd}[column sep=2cm]
B_S(\chi_{A,B})\arrow[squiggly]{r}{\tau(\chi_{A,B},\chi_{C,D})}
\arrow[hookrightarrow]{d}{\beta_S(\chi_{A,B})}  
&B_S(\chi_{C,D})\arrow[hookrightarrow]{d}{\beta_S(\chi_{C,D})}  \\ 
P(\chi_{A,B}) \arrow{r}{\tau(\chi_{A,B},\chi_{C,D})}[swap]{\cong}
&P(\chi_{C,D}) 
\end{tikzcd}
\end{equation}
When $C\subseteq A$, we have $B\subseteq D$, so the condition for $S$
becomes $S\subseteq B\cup C$.  Furthermore, if in addition $T=E\cup
F$, we have 
\[\tau(\chi_{A_1,B_1},\chi_{A_2,B_2})\circ\tau(\chi_{A_2,B_2},\chi_{A_3,B_3})=
\tau(\chi_{A_1,B_1},\chi_{A_3,B_3}).\]

When $A\subseteq C\cup D$, he identity~\eqref{eq:chi2} implies that we
have a~simplicial isomorphism
\[\beta_V(\chi_{C.D}):B_V(\chi_{C,D})\underset\cong\longrightarrow P(\chi_{C\sm A,D\sm A}).\] 
Furthermore, when $S\subseteq C$, the identity~\eqref{eq:chi3} implies
that we have a~simplicial isomorphism
\[X_S(\chi_{C.D}):\gamma_S(\chi_{C,D})\underset\cong\longrightarrow P(\chi_{C\sm S,D\cup S}).\] 

\begin{prop}
Assume $A\cup B=C\cup D$ and $V\subseteq A\cup B$, then
the following diagram commutes
\begin{equation} \label{cd:tau}
\begin{tikzcd}[column sep=1.5cm]
P(\chi_{A,B})\arrow[hookleftarrow]{r}{\beta_V(\chi_{A,B})}
\arrow{d}{\cong}[swap]{\tau(\chi_{A,B},\chi_{C,D})}
& B_V(\chi_{A,B})\arrow[squiggly]{r}{\delta_V(\chi_{A,B})}
& P(\chi_{A,B}\sm V)\arrow{d}{\tau(\chi_{A,B}\sm V,\chi_{C,D}\sm V)}[swap]{\cong} \\   
P(\chi_{C,D})\arrow[hookleftarrow]{r}{\beta_V(\chi_{C,D})}
& B_V(\chi_{C,D})\arrow[squiggly]{r}{\delta_V(\chi_{C,D})}
& P(\chi_{C,D}\sm V)
\end{tikzcd}
\end{equation}
\end{prop}
\pr Consider the diagram on Figure~\ref{cdf:1}.
\begin{figure}[hbt]
\begin{tikzcd}[column sep=1.5cm]
P(\chi_{A,B})\arrow[hookleftarrow]{r}{\beta_V(\chi_{A,B})}
\arrow{d}{\cong}[swap]{\tau_{A,B}}
& B_V(\chi_{A,B})\arrow[squiggly]{r}{\delta_V(\chi_{A,B})}
& P(\chi_{A,B}\sm V)\arrow{ld}{\tau_{A\sm V,B\sm V}}[swap]{\cong} \\  
\da^{A\cup B}\arrow[hookleftarrow]{r}\arrow[leftarrow]{d}{\cong}[swap]{\tau_{C,D}} 
&\da^{A\cup B\sm V}\arrow[leftarrow]{rd}{\tau_{C\sm V,D\sm V}}[swap]{\cong} \\ 
P(\chi_{C,D})\arrow[hookleftarrow]{r}{\beta_V(\chi_{C,D})}
& B_V(\chi_{C,D})\arrow[squiggly]{r}{\delta_V(\chi_{C,D})}
& P(\chi_{C,D}\sm V)
\end{tikzcd}
\caption{}
\label{cdf:1}
\end{figure}
Both the upper and the lower part of this diagram are versions
of~\eqref{cd:tau3}, hence, they commute. Together, they form the
diagram~\eqref{cd:tau}. \qed


\section{Topology of the immediate snapshot complexes}

\subsection{Immediate snapshot complexes are collapsible pseudomanifolds} $\,$

\nin Before proceeding to the main result, that the immediate snapshot
complexes are simplicially homeomorhic simplices, we give short proofs of the facts that
these complexes are pseudomanifolds, that they are contractible
topological spaces, and stronger, that they are collapsible simplicial
complexes.

We start by showing that $P(\tr)$ is a pseudomanifold.

\begin{df}
Let $K$ be a pure simplicial complex of dimension $n$. Two
$n$-simplices of $K$ are said to be {\bf strongly connected} if there
is a~sequence of $n$-simplices so that each pair of consecutive
simplices has a~common $(n-1)$-dimensional face. The complex $K$ is
said to be {\bf strongly connected} if any two $n$-simplices of $K$
are strongly connected.
\end{df}

Clearly, being strongly connected is an equivalence relation on the
set of all $n$-simplices.

\begin{prop}\label{prop:strc}
For an arbitrary round counter $\tr$, the simplicial complex $P(\tr)$
is strongly connected.
\end{prop}
\pr Set $n:=|\supp\tr|-1$. Proposition~\ref{prop:pure} says that
$P(\tr)$ is a pure simplicial complex of dimension~$n$. We now use
induction on $|\tr|$. If $|\tr|=0$, or more generally, if
$|\act\tr|\leq 1$, then $P(\tr)$ is just a~single simplex,
so it is trivially strongly connected. 

Assume $|\act\tr|\geq 2$, and consider the canonical decomposition of
$P(\tr)$.  By Proposition~\ref{prop:strata}, the simplicial complex
$X_S(\tr)$ is isomorphic to $P(\tr_S)$, for all
$S\subseteq\act\tr$. Since $|\tr_S|=|\tr|-|S|<|\tr|$, and
$\supp\tr_S=\supp\tr$, we conclude that $X_S(\tr)$ is a~pure
simplicial complex of dimension~$n$, which is strongly connected by
the induction assumtion. Thus, any pair of $n$-simplices belonging to
the same subcomplex $X_S(\tr)$ is strongly connected.

Pick now any $p\in\act\tr$, and any $S\subseteq\act\tr$, such that
$p\in S$, and consider the subcomplex $X_{S,p}(\tr)=X_S(\tr)\cap
X_{p}(\tr)$. According to Proposition~\ref{prop:strata}, this
subcomplex is isomorphic to $P(\tr_{S,p})$, in particular, it is
non-empty. Take any $(n-1)$-simplex $\tau$ in $X_{S,p}(\tr)$.  By
induction assumptions for $X_S(\tr)$ and $X_{p}(\tr)$, there exist
$n$-simplices $\sigma_1\in X_S(\tr)$, and $\sigma_2\in
X_{p}(\tr)$, such that $\tau\in\partial\sigma_1$ and
$\tau\in\partial\sigma_2$.  This means, that $\sigma_1$ and $\sigma_2$
are strongly connected.  Since being strongly connected is an
equivalence relation, any two $n$-simplices from $X_S(\tr)$ and
$X_{p}(\tr)$ are strongly connected. This includes the case
$S=\act\tr$, implying that any pair of $n$-simplices in $P(\tr)$ is
strongly connected, so $P(\tr)$ itself is strongly connected. 
\qed

\begin{df}
We say that a~strongly connected pure simplicial complex $K$ is a~{\bf
  pseudomanifold} if each $(n-1)$-simplex of $K$ is a~face of
precisely one or two $n$-simplices of $K$. The $(n-1)$-simplices of
$K$ which are faces of precisely one $n$-simplex of $K$ form
a~simplicial subcomplex of $K$, called the {\bf boundary} of $K$, and
denoted $\partial K$.
\end{df}

\begin{prop}\label{prop:pseudo}
For an arbitrary round counter $\tr$, the simplicial complex $P(\tr)$
is a~psedomanifold, such that $\partial P(\tr)=
\cup_{p\in\supp\tr}B_p(\tr)$, i.e., the subcomplex $\partial P(\tr)$
consists of all simplices $\sigma=((W_0,G_0),\dots,(W_t,G_t))$, such
that $G_0\neq\es$.
\end{prop}
\pr By Proposition~\ref{prop:strc} we already know that $P(\tr)$ is
strongly connected. Set again $n:=|\supp\tr|-1$, and let 
$\tau=((W_0,G_0),\dots,(W_t,G_t))$ be an arbitrary $(n-1)$-simplex of $P(\tr)$.
Note that $\codim\tau=|G_0|+\dots+|G_t|$, hence $\codim\tau=1$ implies that
there exist $0\leq k\leq t$, and $p\in\supp\tr$, such that 
\[G_i=\begin{cases}
\{p\},& \textrm{ if } i=k;\\
\es,  & \textrm{ if } i\neq k.
\end{cases}\]
Set $m:=r(p)+1\|M(p,\sigma)|$. Consider
\[\sigma_1=(W_0,\dots,W_k-1,W_k\cup\{p\},W_{k+1},\dots,W_t,
\underbrace{p,\dots,p}_m),\]  
and if $k\geq 1$, consider also
\[\sigma_2=(W_0,\dots,W_k-1,p,W_k,\dots,W_t,
\underbrace{p,\dots,p}_m).\] Obviously,
$\Gamma(\sigma_1,p)=\Gamma(\sigma_2,p)=\tau$, so
$\tau\in\partial\sigma_1$ and $\tau\in\partial\sigma_2$. Furthermore,
the definition of the ghosting construction implies that these are the
only options to find $\sigma$, such that $\Gamma(\sigma,p)=\tau$.

We conclude that $P(\tr)$ is a~pseudomanifold, whose boundary is a
union of the $(n-1)$ simplices $\tau=((W_0,G_0),\dots,(W_t,G_t))$,
such that $W_0\neq\es$, so then the subcomplex $\partial P(\tr)$
consists of all simplices $\sigma=((W_0,G_0),\dots,(W_t,G_t))$, such
that $G_0\neq\es$.  
\qed

\vskip5pt

Consider now a~quite general situation, where $X$ is an arbitrary
topological space, and $\{X_i\}_{i\in I}$ is a~finite family of
subspace of $X$ covering $X$, that is $I$ is finite and $X=\cup_{i\in
  I}X_i$.

\begin{df}{\rm(\cite[Definition 15.14]{book}).}
The {\bf nerve complex} $\cn$ of a covering $\{X_i\}_{i\in I}$ is
a~simplicial complex whose vertices are indexed by $I$, and a~subset
of vertices $J\subseteq$ spans a~simplex if and only if the
intersection $\cap_{i\in J}X_i$ is not empty.
\end{df}

The nerve complex can be useful because of the following fact.
\begin{lm} \label{lm:nerve}
{\rm(Nerve Lemma, \cite[Theorem 15.21, Remark 15.22]{book}).}  Assume
$K$ is a simplicial complex, covered by a~family of subcomplexes
$\ck=\{K_i\}_{i\in I}$, such that $\cap_{i\in J}K_i$ is empty or
contractible for all $J\subseteq I$, then $K$ is homotopy equivalent
to the nerve complex $\cn(\ck)$.
\end{lm}

\begin{crl}
For an arbitrary round counter $\tr$, the simplicial complex $P(\tr)$
is contractible.
\end{crl}
\pr We use induction on $|\tr|$. If $|\tr|=0$, then $P(\tr)$ is just a
simplex, hence contractible. We assume that $|\tr|\geq 1$, and view
the canonical decomposition $P(\tr)=\cup_{S\subseteq\act\tr} X_S(\tr)$
as a~covering of $P(\tr)$.  By Proposition~\ref{prop:strata},
Corollary~\ref{crl:inter}, and the induction assumption, all the
intersections of the subcomplexes $X_S(\tr)$ with each other are
either empty or contractible. This means, that we can apply the Nerve
Lemma~\ref{lm:nerve}, with $K=P(\tr)$, $I=2^{\act\tr}\sm\{\es\}$, and
$K_i$'s are $X_S(\tr)$'s.

Now, by Corollary~\ref{crl:inter} we see that $X_{\act\tr}\cap
X_S=X_{\act\tr,S}\neq\es$ for all $S\subset\act\tr$. It follows that
the nerve complex of this decomposition as a~cone with apex at
$\act\tr\in I$. Since the nerve complex is contractible, it follows
from the Nerve Lemma~\ref{lm:nerve} that $P(\tr)$ is contractible as
well.  \qed

While contractibility is a property of topological spaces, there is
a stronger combinatorial property called {\it collapsibility} which 
some simplicial complexes may have.

\begin{df}
Let $K$ be a simplicial complex. A pair of simplices $(\sigma,\tau)$ of $K$
is called an~{\bf elementary collapse} if the following conditions are satisfied:
\begin{itemize}
\item $\tau$ is a maximal simplex,
\item $\tau$ is the only simplex which properly contains $\sigma$.
\end{itemize}
A finite simplicial complex $K$ is called {\bf collapsible}, if there
exists a sequence $(\sigma_1,\tau_1),\dots,\allowbreak
(\sigma_t,\tau_t)$ of pairs of simplices of $K$, such that
\begin{itemize}
\item this sequence yields a perfect matching on the set of all simplices of~$K$,
\item for every $1\leq k\leq t$, the pair $(\sigma_k,\tau_k)$ is an elementary collapse
in $K\sm\{\sigma_1,\dots,\sigma_{k-1},\allowbreak\tau_1,\dots,\tau_{k-1}\}$.
\end{itemize}
\end{df}

When $(\sigma,\tau)$ is an elementary collapse, we also say that $\sigma$
is a~{\it free} simplex.

We have shown in Proposition~\ref{prop:pseudo} that for any round counter $\tr$ 
the simplicial complex $P(\tr)$ is a pseudomanifold with boundary $\partial P(\tr)$.
Set 
\[\inte P(\tr):=\bigcup_{\sigma\in P(\tr),\,\,\sigma\notin\partial P(\tr)}\inte\sigma,\]
and, for all $A\subseteq S\subseteq\act\tr$, set
\[\partial X_{S,A}(\tr):=\gamma_{S,A}(\tr)^{-1}(\partial P(\tr_{S,A})),\quad
\inte X_{S,A}(\tr):=\gamma_{S,A}(\tr)^{-1}(\inte P(\tr_{S,A})).\]

\begin{prop}\label{prop:6.9}
Assume $\tr$ is an arbitrary round counter, $A\subset S
\subseteq\act\tr$, and $V\subseteq\supp\tr\sm S$. The simplicial
complex $\partial X_{S,A,V}(\tr)$ is the subcomplex of
$X_{S,A,V}(\tr)$ consisting of all simplices
\[\sigma=((W_0,V),\allowbreak (S\sm A,A),\dots,(W_t,G_t)).\]
\end{prop}
\pr Pick $\sigma\in X_{S,A,V}$, and set $\rho$ to be the composition
of the simplicial isomorphisms $X_{S,A,V}(\tr)\ra B_V(\tr_{S,A})\ra
P(\tr_{S\cup V, A\cup V})$ from the commutative
diagram~\eqref{eq:bar}. 

Assume first that $W_1\cup G_1=S$, then
\[\rho(\sigma)=((W_0\sm G_1,(G_0\cup G_1)\sm(A\cup V)),(W_2,G_2),
\dots,(W_t,G_t)).\]
Clearly $\rho(\sigma)\notin\partial P(\tr_{S\cup V,A\cup V})$ if and
only if $(G_0\cup G_1)\sm(A\cup V)=\es$, i.e., $G_0\cup G_1\subseteq
A\cup V$. Since we know that $A\subseteq G_1$, $V\subseteq G_0$, this
means that $G_0=V$ and $G_1=A$, which implies $W_1=S\sm A$. 

Assume now that $S\subseteq G_1$, then we have 
\[\rho(\sigma)=((W_0\sm S,(G_0\cup S)\sm(A\cup V)),(W_1,G_1\sm S),
(W_2,G_2),\dots,(W_t,G_t)).\] Here we have $\rho(\sigma)\notin\partial
P(\tr_{S\cup V,A\cup V})$ if and only if $(G_0\cup S)\sm(A\cup V)
=\es$, which is impossible, since $V\cap S=\es$, and $A\subset V$.
\qed

\begin{crl}\label{crl:strata}
The simplicial complex $P(\tr)$ can be decomposed as a disjoint union
of the simplex $\da^{\pass\tr}=((\pass\tr,\act\tr))$, and the sets
$\inte X_{S,A,V}$, where $(S,A,V)$ range over all triples satisfying
$A\subset S\subseteq\act\tr$ and $V\subseteq\supp\tr\sm S$.

Specifically, for a simplex $\sigma\in P(\tr)$,
$\sigma=((W_0,G_0),\dots,(W_t,G_t))$, we have: if $t=0$, then
$\sigma\subseteq\Delta^{\pass\tr}$, else $\inte\sigma\subseteq\inte
X_{W_1\sm G_1,G_1,G_0}$.
\end{crl}
\pr Immediate from Proposition~\ref{prop:6.9}.
\qed

\begin{lm}\label{lm:coll}
Assume $\tr$ is a round counter, and $p\in\supp\tr$, then there exists
a sequence of elementray collapses reducing the simplicial complex
$P(\tr)$ to the subcomplex $(\partial P(\tr))\sm\inte B_p(\tr)$.
\end{lm}
\pr The proof is again by induction on $|r|$. The case $|r|=0$ is
trivial. The simplices we need to collapse are precisely those, whose
interior lies in $\inte P(\tr)\cup\inte B_p(\tr)$. Let $\Sigma$ denote
the set of all strata $X_{S,A}$, where $A\subset S\subseteq\act\tr$,
together with all strata $X_{S,A,p}$, where $A\subset
S\subseteq\act\tr$, $p\notin S$. By Corollary~\ref{crl:strata}, the
union of the interiors of the strata in $\Sigma$ is precisely $\inte
P(\tr)\cup\inte B_p(\tr)$.

We describe our collapsing as a~sequence of steps. At each step we
pick a~certain pair of strata $(Y,X)$, where $Y\subset X$, which we
must ``collapse''. Then, we use one of the previous results to show
that as a~simplicial pair $(Y,X)$ is isomorphic to
$(B_t(\tr'),P(\tr'))$, for some round counter $\tr'$, such that
$|\tr'|<|\tr|$. By induction assumption this means that there is
a~sequence of simplicial collapses which removes $\inte X\cup\inte Y$.
Finally, we order these pairs of strata with disjoint interiors
$(Y_1,X_1),\dots,(Y_d,X_d)$ such that for every $1\leq i\leq d$, every
simplex $\sigma\in P(\tr)$, such that $\inte\sigma\subseteq\inte
X_i\cup\inte Y_i$, and every $\tau\supset\sigma$, such that
$\dim\tau=\dim\sigma+1$, we have
\begin{equation}\label{eq:intt}
\inte\tau\subseteq\inte X_1\cup\dots\cup\inte X_i\cup
\inte Y_1\cup\dots\cup\inte Y_i.
\end{equation}
This means, that at step $i$ we can
collapse away the pair of strata $(Y_i,X_i)$ (i.e., collapse away
those simplices whose interior is contained in $\inte X_i\cup\inte
Y_i$) using the procedure given by the induction assumption, and that
these elementary collapses will be legal in $P(\tr)\sm
(X_1\cup\dots\cup\inte X_{i-1}\cup\inte Y_1\cup\dots\cup\inte
Y_{i-1})$ as well.

Our procedure is now divided into $3$ stages.  At stage 1, we match
the strata $X_{S,A,p}$ with $X_{S,A}$, for all $A\subset S
\subseteq\act\tr$, such that $p\notin S$. It follows from the
commutativity of the diagram~\eqref{eq:bar} that each pair of
simplicial subcomplexes $(X_{S,A,p},X_{S,A})$ is isomorphic to the
pair $(B_p(\tr_{S,A}),P(\tr_{S,A}))$. We have
$|\tr_{S,A}|\leq|\tr|-|S|<|\tr|$, hence by induction assumption, this
pair can be collapsed. As a collapsing order we choose any order which
does not decrease the cardinality of the set $A$. Take $\sigma$ such
that $\inte\sigma\subseteq\inte X_{S,A,p}\cup\inte X_{S,A}$. By
Proposition~\eqref{prop:6.9} this means that $\sigma=((W_0,T),(S\sm A,A),\dots)$,
where either $T=\es$, or $T=\{p\}$. Take $\tau\supset\sigma$,
such that $\dim\tau=\dim\sigma+1$. Then by Proposition~\ref{prop:b}(b) there exists
$q\in A(\tau)$, such that $\sigma=\Gamma_q(\tau)$. 
A~case-by-case analysis of the ghosting construction shows that 
$\inte\tau\subseteq\inte X$, where $X$ is one of the following strata:
$X_{S,A}$, $X_{S,A,p}$, $X_q$, $X_{q,\es,p}$, $X_{S,A\sm\{q\}}$, 
$X_{S,A\sm\{q\},p}$. Since the order in which we do collapses
does not decrease the cardinality of $A$, the interiors of the last $4$ of 
these strata have already been removed, hence the condition~\eqref{eq:intt}
is satisfied.

At stage 2, we match $X_S$ with $X_{S,S\sm\{p\}}$, for all
$S\subseteq\act\tr$, such that $p\in S$, $|S|\geq 2$. By commutativity of the 
diagram~\eqref{cd:b1}, the pair $(X_{S,S\sm\{p\}},X_S)$ is isomorphic to
$(B_{S\sm\{p\}}(\tr_S),P(\tr_S))$. This big collapse can easily be
expressed as a sequence of elementary collapses, though in
a~non-canonical way.  For this, we pick any $q\in S\sm\{p\}$. It
exists, since we assumed that $|S|\geq 2$.  Then we match pairs
$(X_{S,A\cup\{q\}},X_{S,A})$, for all $A\subseteq S\sm\{q\}$.  
Again, by commutativity of the diagram~\eqref{cd:b1}, this pair 
is isomorphic to $(B_q(\tr_{S,A}),P(\tr_{S,A}))$. The
order in which we arrange $S$ does not matter for the collapsing
order.  Once $S$ is fixed, the collapsing order inside does not
decrease the cardinality of $A$. As above, take $\sigma$ such
that $\inte\sigma\subseteq\inte X_{S,A\cup\{q\}}\cup\inte X_{S,A}$,
take $\tau\supset\sigma$, such that $\dim\tau=\dim\sigma+1$, and take
$r\in A(\tau)$, such that $\sigma=\Gamma_q(\tau)$. By
Proposition~\eqref{prop:6.9} we have $\sigma=((W_0,\es),(S\sm A,A),\dots)$,
or $\sigma=((W_0,\es),(S\sm(A\cup\{q\}),A\cup\{q\}),\dots)$. Note,
that both $q$ and $r$ are different from $p$, but we may have $q=r$.
Again, a~case-by-case analysis of the ghosting construction shows that 
$\inte\tau\subseteq\inte X$, where $X$ is one of the following strata:
$X_{S,A}$, $X_{S,A\cup\{q\}}$, $X_{S,A\sm\{r\}}$, $X_{S,A\cup\{q\}\sm\{r\}}$, 
$X_q$, $X_r$. Again, since collapsing order does not decrease the cardinality
of~$A$, the condition~\eqref{eq:intt} is satisfied.

At stage 3, we collapse the pair $(X_{p,p},X_p)$. Let us be specific. 
First, by Corollary~\ref{crl:xaa} we know that 
$X_{p,p}=\bigcup_{\{p\}\subset S\subseteq\act\tr}X_{S,p}$, and 
it follows from Proposition~{prop:6.9} that 
$\inte X_{p,p}=\bigcup_{\{p\}\subset S\subseteq\act\tr}\inte X_{S,p}$. 
By commutativity of the diagram~\eqref{cd:b1}, the pair $(X_{p,p},X_p)$
is isomorphic to $(B_p(\tr_p),P(\tr_p))$, hence it can be collapsed
using the induction assumption. Clearly, the entire procedure exhausts
the set $\Sigma$, and we arrive at the simplicial complex 
$(\partial P(\tr))\sm\inte B_p(\tr)$.
\qed

\begin{crl}
For an arbitrary round counter $\tr$, the simplicial complex $P(\tr)$
is collapsible.
\end{crl}
\pr Iterative use of Lemma~\ref{lm:coll}. 
\qed

\subsection{Homeomorphic gluing}

\begin{df}
We say that a simplicial complex $K$ is {\bf simplicially homeomorphic}
to a~simplex $\Delta^A$, where $A$ is some finite set, if there exists 
a~homeomorphism $\varphi:\da^A\ra K$, such that for every simplex 
$\sigma\in\da^A$, the image $\varphi(\sigma)$ is a~subcomplex of~$K$.
\end{df}

When we say that a CW complex is {\it finite} we shall mean that it has
finitely many cells.

\begin{df}\label{df:gd}
 Let $X$ and $Y$ be finite CW complexes. A~{\bf homeomorphic gluing
   data} between $X$ and $Y$ consists of the following:
\begin{itemize}
\item a~family $(A_i)_{i=1}^t$ of CW subcomplexes of $X$, such that
  $X=\cup_{i=1}^t A_i$,
\item a~family $(B_i)_{i=1}^t$ of CW subcomplexes of $Y$, such that
  $Y=\cup_{i=1}^t B_i$,
\item a~family of homeomorphisms $(\varphi_i)_{i=1}^t$,
  $\varphi_i:A_i\rightarrow B_i$,
\end{itemize}
satisfying the compatibility condition: if $x\in A_i\cap A_j$, then
$\varphi_i(x)=\varphi_j(x)$.
\end{df}

Given finite CW complexes $X$ and $Y$, together with homeomorphic
gluing data $(A_i,B_i,\varphi_i)_{i=1}^t$ from $X$ to $Y$, we define
$\varphi:X\rightarrow Y$, by setting $\varphi(x):=\varphi_i(x)$,
whenever $x\in A_i$. The compatibility condition from
Definition~\ref{df:gd} implies that $\varphi(x)$ is independent of the
choice of $i$, hence the map $\varphi:X\rightarrow Y$ is well-defined.

\begin{lm} \label{lm:hg} {\rm (Homeomorphism Gluing Lemma).}

\noindent Assume we are given finite CW complexes $X$ and $Y$, and
homeomorphic gluing data $(A_i,B_i,\varphi_i)_{i=1}^t$, satisfying an
additional condition:
\begin{equation} \label{eq:c2}
\textrm{ if } \varphi(x)\in B_i, \textrm{ then } x\in A_i,
\end{equation}
then the map $\varphi:X\rightarrow Y$ is a~homeomorphism.
\end{lm}
\pr First it is easy to see that $\varphi$ is surjective. 
Take an arbitrary $y\in Y$, then there exists $i$ such that $y\in B_i$.
Take $x=\varphi_i^{-1}(y)$, clearly $\varphi(x)=y$.

Let us now check the injectivity of $\varphi$. Take $x_1,x_2\in X$
such that $\varphi(x_1)=\varphi(x_2)$. There exists $i$ such that
$x_1\in A_i$.  Then $\varphi(x_1)=\varphi_i(x_1)\in B_i$, hence
$\varphi(x_2)\in B_i$.  Condition~\eqref{eq:c2} implies that $x_2\in
A_i$. The fact that $x_1=x_2$ now follows from the injectivity of
$\varphi_i$.

We have verified that $\varphi$ is bijective, so
$\varphi^{-1}:Y\rightarrow X$ is a well-defined map. We shall now
prove that $\varphi^{-1}$ is continuous by showing that $\varphi$
takes closed sets to closed sets. To start with, let us recall the
following basic property of the topology of CW complexes: a subset $A$
of a CW complex $X$ is closed if and only if its intersection with the
closure of each cell in $X$ is closed. Sometimes, one uses the
terminology {\it weak topology} of the CW complex. This property was
an integral part of the original J.H.C.\ Whitehead definition of CW
complexes, see, e.g.,~\cite[Proposition A.2.]{Hat} for further
details.

Let us return to our situation. We claim that $A\subseteq X$ is
closed, if and only if $A\cap A_i$ is closed in $A_i$, for each
$i=1,\dots,t$. Note first that since $A_i$ is itself closed, a~subset
$S\subseteq A_i$ is closed in $X$ if and only if it is closed in
$A_i$, so we will skip mentioning where the sets are closed.  Clearly,
if $A$ is closed, then $A\cap A_i$ is closed for all~$i=1,\dots,t$.
On the other hand, assume $A\cap A_i$ is closed for all~$i$. Let
$\sigma$ be a closed cell of $X$, we need to show that $A\cap\sigma$
is closed.  Since $X=\cup_{i=1}^t A_i$, and $A_i$'s are CW
subcomplexes of $X$, there exists $i$, such that $\sigma\subseteq
A_i$. Then $A\cap\sigma=A\cap (A_i\cap\sigma)=(A\cap A_i)\cap\sigma$,
but $(A\cap A_i)\cap\sigma$ is closed since $A\cap A_i$ is closed.
Hence $A\cap\sigma$ is closed and our argument is finished.
Similarly, we can show that $B\subseteq X$ is closed, if and only if
$B\cap B_i$ is closed, for each $i=1,\dots,t$.
 
Pick now a~closed set $A\subseteq X$, we want to show that
$\varphi(A)$ is closed.  To start with, for all $i$ the set $A\cap
A_i$ is closed, hence $\varphi_i(A\cap A_i)\subseteq B_i$ is also
closed, since $\varphi_i$ is a~homeomorphism.  Let us verify that for
all $i$ we have
\begin{equation}\label{eq:jhc}
\varphi_i(A\cap A_i)=\varphi(A)\cap B_i.
\end{equation} 
Assume $y\in\varphi_i(A\cap A_i)$. On one hand $y\in\varphi_i(A_i)$,
so $y\in B_i$, on the other hand, $y=\varphi_i(x)$, for $x\in A$, so
$y\in\varphi(A)$. Reversely, assume $y\in\varphi(A)$ and $y\in
B_i$. Then $y=\varphi(x)\in B_i$, so condition~\eqref{eq:c2} implies
that $x\in A_i$, hence $y\in\varphi(A\cap A_i)$, which
proves~\eqref{eq:jhc}. It follows that $\varphi(A)\cap B_i$ is closed
for all $i$, hence $\varphi(A)$ itself is closed. This proves that
$\varphi^{-1}$ is continuous.

We have now shown that $\varphi^{-1}:Y\rightarrow X$ is a~continuous
bijection.  Since $X$ and $Y$ are both finite CW complexes, they are
compact Hausdorff when viewed as topological spaces.  It is a~basic
fact of set-theoretic topology that a~continuous bijection between
compact Hausdorff topological spaces is automatically a~homeomorphism,
see e.g., \cite[Theorem 26.6]{Mun}.  \qed

\vspace{5pt}

The following variations of the Homeomorphism Gluing Lemma~\ref{lm:hg}
will be useful for us. 

\begin{crl} \label{crl:hg}
 Assume we are given finite CW complexes $X$ and $Y$, and
homeomorphic gluing data $(A_i,B_i,\varphi_i)_{i=1}^t$, satisfying an
additional condition:
\begin{equation} \label{eq:c3}
\textrm{ for all } I\subseteq[t]: \varphi:A_I\rightarrow B_I
\textrm{ is a~bijection. }
\end{equation}
Then the map $\varphi:X\rightarrow Y$ is a~homeomorphism.
\end{crl}
\pr Clearly, we just need to show that the condition \eqref{eq:c3}
implies the condition~\eqref{eq:c2}. Assume $y=\varphi(x)$, $y\in
B_i$, and $x\not\in A_i$. Let $I$ be the maximal set such that $y\in
B_I$. The condition~\eqref{eq:c3} implies that there exists a~unique
element $\tilde x\in A_I$, such that $\varphi(\tilde x)=y$. In
particular, $\tilde x\in A_i$, hence $x\neq\tilde x$. Even stronger,
if $x\in A_i$, for some $i\in I$, then $x,\tilde x\in A_i$, hence
$x=\tilde x$, since $\varphi_i$ is injective. So $x_i\not\in A_i$, for
all $i\in I$. Hence, there exists $j\not\in I$, such that $x\in A_j$,
which implies $\varphi(x)\in B_j$, yielding a~contradiction to the
maximality of the set~$I$. \qed

\begin{crl}\label{crl:hg2}
Assume we are given CW complexes $X$ and $Y$, a collection
$(A_i)_{i=1}^t$ of CW subcomplexes of $X$, a collection
$(B_i)_{i=1}^t$ of CW subcomplexes of $Y$, and a collection
$(\varphi_I)_{I\subseteq[t]}$ of maps such that 
\begin{itemize}
\item $X=\cup_{i=1}^t A_i$, $Y=\cup_{i=1}^t B_i$; 
\item for every $I\subseteq[t]$, the map $\varphi_I:A_I\rightarrow
  B_I$ is a~homeomorphism;
\item for every $J\supseteq I$ the following diagram commutes
\begin{equation} \label{cd:ij1}
\begin{tikzcd}
A_J\arrow{r}{\varphi_J}[swap]{\cong}\arrow[hookrightarrow]{d} 
&B_J\arrow[hookrightarrow]{d} \\
A_I \arrow{r}{\varphi_I}[swap]{\cong} &B_I
\end{tikzcd}
\end{equation}
\end{itemize}
Then $(A_i,B_i,\varphi_i)_{i=1}^t$ is a homeomorphic gluing data, and
the map $\varphi:X\rightarrow Y$ defined by this data is
a~homeomorphism.
\end{crl}
\pr For arbitrary $1\leq i,j\leq t$, commutativity of \eqref{cd:ij1}
implies that also the following diagram is commutative
\[\begin{tikzcd}
A_i\arrow[hookleftarrow]{r}\arrow{d}{\varphi_{\{i\}}}[swap]{\cong} 
&A_{\{i,j\}}\arrow[hookrightarrow]{r}\arrow{d}{\varphi_{\{i,j\}}}[swap]{\cong} 
&A_j\arrow{d}{\varphi_{\{j\}}}[swap]{\cong} \\
B_i\arrow[hookleftarrow]{r} 
&B_{\{i,j\}}\arrow[hookrightarrow]{r} &B_j
\end{tikzcd}\]
In other words, for any $x\in A_i\cap A_j$, we have
$\varphi_{\{i\}}(x)=\varphi_{\{i,j\}}(x)=\varphi_{\{j\}}(x)$.  It
follows that $(A_i,B_i,\varphi_{\{i\}})_{i=1}^t$ is a~homemorphic
gluing data. Since for all $I\subseteq[t]$, the map $\varphi_I$ is
a~homeomorphism, it is in particular bijective, so conditions of
Corollary~\ref{crl:hg} are satisfied, and the defined map $\varphi$ is
a~homeomorphism. \qed


\subsection{Main Theorem}

The fact that the protocol complexes in the immediate shapshot
read/write shared memory model are homeomorphic to simplices has
been folklore knowledge in the theoretical distributed computing
community, \cite{Herl}. The next theorem provides a rigorous
mathematical proof of this fact.

\begin{thm}\label{thm:main}

For every round counter $\tr$ there exists a~homeomorphism 
\[\Phi(\tr):P(\tr)\stackrel\cong\longrightarrow P(\chi(\tr)),\]
such that 
\begin{enumerate}
\item[(1)] for all $V\subset\supp\tr$ the following diagram commutes:
\begin{equation}\label{cd:b}
\begin{tikzcd}[column sep=1.1cm]
P(\tr\sm V)\arrow[leftsquigarrow]{r}{\delta_V(\tr)} 
\arrow{d}{\cong}[swap]{\Phi(\tr\sm V)}
&B_V(\tr)\arrow[hookrightarrow]{r}{\beta_V(\tr)} 
& P(\tr)\arrow{d}{\Phi(\tr)}[swap]{\cong} \\ 
P(\chi(\tr\sm V))\arrow[leftsquigarrow]{r}{\delta_V(\chi(\tr))} 
& B_V(\chi(\tr))\arrow[hookrightarrow]{r}{\beta_V(\chi(\tr))} 
&P(\chi(\tr)) 
\end{tikzcd}
\end{equation}
\item[(2)]  for all $S\subseteq\act\tr$ the following diagram commutes:
\begin{equation}\label{cd:b2}
\begin{tikzcd}[column sep=1.1cm]
X_S(\tr)\arrow[squiggly]{r}{\gamma_S(\tr)}\arrow[hookrightarrow]{rd}{\alpha_S(\tr)}
&P(\tr_S)\arrow{r}{\Phi(\tr_S)}[swap]{\cong}
&P(\chi(\tr_S))\arrow{r}{\tau}[swap]{\cong}
&P(\chi(\tr)_S)\arrow[leftsquigarrow]{r}{\gamma_S(\chi(\tr))}
&X_S(\chi(\tr))\\
&P(\tr)\arrow{rr}{\Phi(\tr)}[swap]{\cong}
& & P(\chi(\tr))\arrow[hookleftarrow]{ru}{\alpha_S(\chi(\tr))}
\end{tikzcd}
\end{equation}
where $\tau=\tau(\chi(\tr_S),\chi(\tr)_S)$.
\end{enumerate}
In particular, the complex $P(\tr)$ is simplicially homemorphic to $\da^{\supp\tr}$.
\end{thm}

\pr Our proof is a~double induction, first on $|\supp\tr|$, then, once
$|\supp\tr|$ is fixed, on the cardinality of the round counter
$\tr$. As a~base of the induction, we note that the case
$|\supp\tr|=1$ is trivial, since the involved spaces are
points. Furthermore, if $|\supp\tr|$ is fixed, and $|\tr|=0$, we take
$\Phi(\tr)$ to be the identity map. In this case the simplicial
complexes $P(\tr)$ and $P(\chi(\tr))$ are simplices. The
diagram~\eqref{cd:b} commutes, since also $\Phi(\tr\sm V)$ is the
identity map. The condition (2) of the theorem is void, since
$\act\tr=\es$. As a~matter of fact, more generally, $\Phi(\tr)$ can be
taken to be the identity map whenever $\tr=\chi(\tr)$, that is
whenever $\tr(i)\in\{0,1\}$, for all $i\in\supp\tr$.

We now proceed to prove the induction step, assuming that $|\tr|\geq
1$.  For every pair of sets $A\subseteq S\subseteq\act\tr$, such that
$S\neq\es$, we define a~map
\[\varphi_{S,A}(\tr):X_{S,A}(\tr)\longrightarrow X_{S,A}(\chi(\tr)),\]
as follows
\begin{equation} \label{cd:phi}
\begin{tikzcd}[column sep=0.9cm]
\varphi_{S,A}(\tr):X_{S,A}(\tr)\arrow[squiggly]{r}{\gamma_{S,A}(\tr)} 
&P(\tr_{S,A}) \arrow{r}{\Phi(\tr_{S,A})}[swap]{\cong} 
&P(\chi(\tr_{S,A}))\arrow{r}{\tau}[swap]{\cong} 
&P(\chi(\tr)_{S,A})\arrow[leftsquigarrow]{r}{\gamma_{S,A}(\chi(\tr))} 
&X_{S,A}(\chi(\tr)),
\end{tikzcd}
\end{equation}
where $\tau=\tau(\chi(\tr_{S,A}),\chi(\tr)_{S,A})$. Since
$|\tr_{S,A}|\leq|\tr_S|=|\tr|-|S|<|\tr|$, the map $\Phi(\tr_{S,A})$ is
already defined by induction, so $\varphi_{S,A}(\tr)$ is well-defined
by the sequence~\eqref{cd:phi}. Obviously, the map $\varphi_{S,A}$ is
a~homeomorphism for all pairs $S,A$.

We want to use Corollary~\ref{crl:hg2} to construct the global
homeomorphism $\Phi(\tr)$ by gluing the local ones
$\varphi_{S,A}(\tr)$. In our setting here, the notations of
Corollary~\ref{crl:hg2} translate to $X=P(\tr)$, $Y=P(\chi(\tr))$,
$A_I$'s are $X_{S,A}(\tr)$'s, $B_I$'s are $X_{S,A}(\chi(\tr))$'s, and
$\varphi_I$'s are $\varphi_{S,A}(\tr)$'s. To satisfy the conditions of
Corollary~\ref{crl:hg2}, we need to check that the following diagram
commutes whenever $X_{S,A}\subseteq X_{T,B}$
\begin{equation}\label{cd:m1}
\begin{tikzcd}
X_{S,A}(\tr)\arrow{r}{\varphi_{S,A}(\tr)}[swap]{\cong}
\arrow[hookrightarrow]{d}{i} 
&X_{S,A}(\chi(\tr))\arrow[hookrightarrow]{d}{j}  \\ 
X_{T,B}(\tr)\arrow{r}{\varphi_{T,B}(\tr)}[swap]{\cong} 
&X_{T,B}(\chi(\tr)),  
\end{tikzcd}
\end{equation}
where $i$ and $j$ denote the inclusion maps.

Note, that by Proposition~\ref{pr:inc1}, we have $X_{S,A}\subseteq
X_{T,B}$ if and only if either $S=T$ and $B\subseteq A$, or
$T\subseteq A$.  Consider first the case $S=T$, $B=\es$. Consider the
diagram on Figure~\ref{cdf:m2}.
\begin{figure}[hbt]
\begin{tikzcd}
\,& P(\tr_{S,A})\arrow{r}{\Phi}[swap]{\cong} 
&P(\chi(\tr_{S,A}))\arrow{r}{\tau}[swap]{\cong} 
&P(\chi(\tr)_{S,A}) \\
X_{S,A}(\tr)\arrow[squiggly]{ru}{\gamma}\arrow[hookrightarrow]{d} 
&B_A(\tr_S)\arrow[squiggly]{u}[swap]{\delta}\arrow[hookrightarrow]{d}{\beta} 
&B_A(\chi(\tr_S))\arrow[squiggly]{u}[swap]{\delta}\arrow[hookrightarrow]{d}{\beta} 
&B_A(\chi(\tr)_S)\arrow[hookrightarrow]{d}{\beta}\arrow[squiggly]{u}[swap]{\delta}
&X_{S,A}(\chi(\tr))\arrow{d}\arrow[squiggly]{lu}{\gamma} \\
X_S(\tr)\arrow[squiggly]{r}{\gamma} 
&P(\tr_S)\arrow{r}{\Phi}[swap]{\cong} 
&P(\chi(\tr_S))\arrow{r}{\tau}[swap]{\cong} &P(\chi(\tr)_S) 
&X_S(\chi(\tr))\arrow[squiggly]{l}{\gamma}
\end{tikzcd}
\caption{}
\label{cdf:m2}
\end{figure}
The leftmost pentagon is the diagram \eqref{cd:b1}, which commutes by
Proposition~\ref{prop:5}. The following hexagon is the diagram
\eqref{cd:b}, where $\tr$ is replaced with $\tr_S$. Since
$|\tr_S|=|\tr|-|S|<|\tr|$, this diagram commutes by induction. The
next hexagon is the diagram \eqref{cd:tau}, for
$\chi_{C_1,D_1}=\chi(\tr_S)$, $\chi_{C_2,D_2}=\chi(\tr)_S$, and we use
the fact that $\chi(\tr_S)\sm A=\chi(\tr_{S,A})$. Finally, the
rightmost pentagon is also the commuting diagram \eqref{cd:b1}, where
$\tr$ is replaced with $\chi(\tr)$. Since removing the $3$ inner terms
of the diagram on Figure \ref{cdf:m2} yields the diagram~\eqref{cd:m1} with
$S=T$, $B=\es$, we conclude that \eqref{cd:m1} commutes in this
special case.

Consider now the case $S=T$, $B\subseteq A$. We have inclusions
$X_{S,A}\hra X_{S,B}\hra X_{S}$, and it is easy to see that the
commutativity of the diagram \eqref{cd:m1} for the inclusion
$X_{S,A}\hra X_{S,B}$ follows from the commutativity of the diagrams
\eqref{cd:m1} for the inclusions $X_{S,A}\hra X_S$ and $X_{S,B}\hra
X_S$. Hence we are done with the proof of this case.

Let us now prove the commutativity of the diagram~\eqref{cd:m1} for
the inclusion $X_{S,A}\hra X_{T,B}$, when $T\subseteq A$.
Assume first that $A=T=B\neq\es$, and consider the diagram on Figure~\ref{cdf:mt1}.

\begin{figure}[hbt]
\begin{tikzcd}[column sep=0.6cm]
\, &P(\tr_{S,A})\arrow{r}{\Phi}[swap]{\cong} 
&P(\chi(\tr_{S,A}))\arrow{r}{\tau}[swap]{\cong} 
& P(\chi(\tr)_{S,A}) \\
X_{S,A}(\tr)\arrow[hookrightarrow]{d}\arrow[squiggly]{ru}{\gamma} 
&X_{\wti S}(\tr\sm A)\arrow[hookrightarrow]{d}{\alpha}\arrow[squiggly]{u}[swap]{\gamma}  &  
&X_{\wti S}(\chi(\tr\sm A))\arrow[hookrightarrow]{d}{\alpha}\arrow[squiggly]{u}[swap]{\gamma} 
&X_{S,A}(\chi(\tr))\arrow[hookrightarrow]{d}\arrow[squiggly]{lu}[swap]{\gamma}  \\
X_{A,A}(\tr)\arrow[squiggly]{r}{\gamma} 
&P(\tr\sm A)\arrow{rr}{\Phi}[swap]{\cong}  & 
&P(\chi(\tr\sm A))
&X_{A,A}(\chi(\tr))\arrow[squiggly]{l}[swap]{\gamma},
\end{tikzcd}
\caption{}
\label{cdf:mt1}
\end{figure}
where $\wti S=S\sm A$. A few of the maps in the diagram on Figure~\ref{cdf:mt1} 
need to be articulated.  To start with, we have the identity $\tr_{A,A}=\tr\sm
A$, explaining the simplicial isomorphism
$\gamma_{A,A}(\tr):X_{A,A}(\tr)\ra P(\tr\sm A)$. Similarly,
$\chi(\tr)_{A,A}=\chi(\tr\sm A)$ explains
$\gamma_{A,A}(\chi(\tr)):X_{A,A}(\chi(\tr))\ra P(\chi(\tr\sm A))$.
Furthermore, by~\eqref{eq:rsa1} we have $\tr\sm A\dar\wti S=\tr\dar
S\sm A=\tr_{S,A}$ and $\chi(\tr\sm A)\dar\wti S=\chi(\tr)\sm A\dar\wti
S=\chi(\tr)_{S,A}$. These identities explain the presence of the maps
$\gamma_{\wti S}(\tr\sm A):X_{\wti S}(\tr\sm A)\ra P(\tr_{S,A})$, and
$\gamma_{\wti S}(\chi(\tr\sm A)):X_{\wti S}(\chi(\tr\sm A))\ra
P(\chi(\tr)_{S,A})$.

Let us look at the commutativity of the diagram on Figure~\ref{cdf:mt1}. The middle
heptagon is the diagram~\eqref{cd:b2} with $\tr\sm A$ instead of $\tr$
and $\wti S$ instead of $S$; where we again use the identity $\tr\sm
A\dar\wti S=\tr\dar S\sm A$. Since $|\supp(\tr\sm
A)|=|\supp\tr|-|A|<|\supp\tr|$, the induction hypothesis implies that
this heptagon commutes. The leftmost pentagon is \eqref{cd:xs}, with
$\wti S$ instead of $S$, whereas the rightmost pentagon is
\eqref{cd:xs} as well, this time with $\wti S$ instead of $S$, and
$\chi(\tr)$ instead of~$\tr$. They both commute by
Proposition~\ref{prop:cxs}. Again, removing the $2$ inner terms from
the diagram on Figure~\ref{cdf:mt1} will yield the diagram~\eqref{cd:m1} with
$A=T=B$, so we conclude that \eqref{cd:m1} commutes in this special
case.

In general, when $T\subseteq A$, we have a sequence of inclusions
$X_{S,A}\hra X_{S,T}\hra X_{T,T}\hra X_{T,B}$. Again, it is easy to
see that the commutativity of the diagram \eqref{cd:m1} for the
inclusion $X_{S,A}\hra X_{T,B}$ follows from the commutativity of the
diagrams \eqref{cd:m1} for the inclusions $X_{S,A}\hra X_{S,T}$,
$X_{S,T}\hra X_{T,T}$, and $X_{T,T}\hra X_{T,B}$. Hence we are done
with the proof of this case as well.


We now know that $\Phi(\tr)$ is a well-defined homeomorphism between
$P(\tr)$ and $P(\chi(\tr))$. To finish the proof of the main theorem,
we need to check the commutativity of the diagrams~\eqref{cd:b}
and~\eqref{cd:b2}. The commutativity of~\eqref{cd:b2} is an~immediate
consequence of~\eqref{cd:phi}, and the way $\Phi(\tr)$ was defined. To
show that~\eqref{cd:b} commutes, pick any $S\subseteq\act\tr$, which
is disjoint from $A$, and consider the diagram on Figure~\ref{cd:mt2}.
\begin{figure}[hbt]
\begin{tikzcd}
P(\chi(\tr))\arrow[hookleftarrow]{rr}{\alpha}\arrow[hookleftarrow]{dd}{\beta} & 
&X_S(\chi(\tr))\arrow{rr}{\rho}[swap]{\cong} 
& &P(\chi(\tr_S))\arrow[hookleftarrow]{dd}{\beta} \\
&P(\tr)\arrow[hookleftarrow]{d}{\beta}\arrow{ul}{\cong}[swap]{\Phi}\arrow[hookleftarrow]{r}{\alpha}
&X_S(\tr)\arrow[squiggly]{r} \arrow[hookleftarrow]{d}
&P(\tr_S)\arrow{ur}{\Phi}[swap]{\cong}\arrow[hookleftarrow]{d}{\beta} \\
B_V(\chi(\tr))\arrow[squiggly]{dd}{\delta} 
&B_V(\tr)\arrow[squiggly]{d}{\delta}\arrow[hookleftarrow]{r}
&X_{S,\es,V}(\tr)\arrow[squiggly]{d}{\psi}\arrow[squiggly]{r}{\varphi} 
&B_V(\tr_S)\arrow[squiggly]{d}{\delta} 
&B_V(\chi(\tr_S))\arrow[squiggly]{dd}{\delta}\\
&P(\tr\sm V)\arrow{dl}{\cong}[swap]{\Phi}\arrow[hookleftarrow]{r}{\alpha}
&X_S(\tr\sm V)\arrow[squiggly]{r} 
&P(\bar r_{S,V})\arrow{dr}{\Phi}[swap]{\cong} \\
P(\chi(\tr\sm V))\arrow[hookleftarrow]{rr}{\alpha}& 
&X_S(\chi(\tr\sm V))\arrow{rr}{\nu}[swap]{\cong} & 
&P(\chi(\bar r_{S,V}))
\end{tikzcd}
\caption{}
\label{cd:mt2}
\end{figure}
The maps $\varphi$ and $\psi$ are as in Proposition~\ref{prop:6.13},
and the maps $\rho$ and $\nu$ are given by
\[
\begin{tikzcd}[column sep=2cm]
\rho:X_S(\chi(\tr))\arrow[squiggly]{r}{\gamma_S(\chi(\tr))}
&P(\chi(\tr)_S)\arrow{r}{\tau(\chi(\tr_S),\chi(\tr)_S)^{-1}}
&P(\chi(\tr_S))
\end{tikzcd}
\] 
and
\[
\begin{tikzcd}[column sep=2cm]
\nu:X_S(\chi(\tr\sm A))\arrow[squiggly]{r}{\gamma_S(\chi(\tr\sm A))}
&P(\chi(\tr)_{S,A})\arrow{r}{\tau(\chi(\tr_{S,A}),\chi(\tr)_{S,A})^{-1}}
&P(\chi(\tr_{S,A})),
\end{tikzcd}
\] 
where we use the identities $\chi(\tr\sm A)=\chi(\tr)\sm A$,
$\chi(\tr_S)\sm A=\chi(\tr_{S,A})$, and $\chi(\tr\sm A)_S=\chi(\tr)_{S,A}$,
with the latter one relying on the fact that $S\cup A=\es$.

Let us investigate the diagram on Figure~\ref{cd:mt2} in some detail. The
middle part is precisely the diagram~\eqref{eq:bar}, which commutes by
Proposition~\ref{prop:6.13}. We have $4$ hexagons surround that middle
part.  The hexagon on the left is the diagram~\eqref{cd:b} itself. The
hexagon above is precisely the diagram~\eqref{cd:b2}, so it
commutes. The hexagon below is the diagram~\eqref{cd:b2} with $\tr\sm
A$ instead of $\tr$, where we use \eqref{eq:rsa2} again. This diagram
commutes by the induction hypothesis. The hexagon on the right is the
diagram~\eqref{cd:b} with $\tr_S$ instead of $\tr$. Since
$|\tr_S|<|\tr|$, it also commutes by the induction assumption.

Let us now show that the diagram obtained from the one on Figure~\ref{cd:mt2} 
by the removal of the $9$ inner terms commutes. This diagram can be factorized 
as shown on Figure~\ref{cdf:mt3}.
\begin{figure}
\begin{tikzcd}
P(\chi(\tr))\arrow[hookleftarrow]{r}{\alpha}\arrow[hookleftarrow]{d}{\beta} 
&X_S(\chi(\tr))\arrow[squiggly]{r}{\gamma} 
&P(\chi(\tr)_S)\arrow[leftarrow]{r}{\tau}[swap]{\cong}\arrow[hookleftarrow]{d}{\beta} 
&P(\chi(\tr_S))\arrow[hookleftarrow]{d}{\beta} \\
B_V(\chi(\tr))\arrow[squiggly]{d}{\delta} & 
&B_V(\chi(\tr)_s)\arrow[squiggly]{d}{\delta} 
&B_V(\chi(\tr_S))\arrow[squiggly]{d}{\delta}\\
P(\chi(\tr\sm V))\arrow[hookleftarrow]{r}{\alpha} 
&X_S(\chi(\tr\sm V))\arrow[squiggly]{r}{\gamma} 
&P(\chi(\tr)_{S,V})\arrow[leftarrow]{r}{\tau}[swap]{\cong}
&P(\chi(\tr_{S,V}))
\end{tikzcd}
\caption{}
\label{cdf:mt3}
\end{figure}
The left part of the diagram on Figure~\ref{cdf:mt3} is \eqref{eq:bar} with
$\chi(\tr)$ instead of $\tr$, whereas the right part of the diagram on 
Figure~\ref{cdf:mt3} is the diagram~\eqref{cd:tau} with $\chi_{C_1,D_1}=\chi(\tr_S)$,
$\chi_{C_2,D_2}=\chi(\tr)_S$. They both commute, hence so does the
whole diagram.

Consider now two sequences of maps in the diagram on Figure~\ref{cd:mt2}:
\begin{equation}
\label{eq:s1}
\begin{tikzcd}[column sep=0.6cm]
B_V(\tr)\cap X_S(\tr)\arrow[hookrightarrow]{r}
&B_V(\tr)\arrow[hookrightarrow]{r}{\beta}
&P(\tr)\arrow{r}{\Phi}[swap]{\cong}&P(\chi(\tr))
\end{tikzcd}
\end{equation}
and
\begin{equation}
\label{eq:s2}
\begin{tikzcd}[column sep=0.5cm]
X_{S,\es,V}(\tr)\arrow[hookrightarrow]{r}
&B_V(\tr)\arrow[squiggly]{r}{\delta}
&P(\tr\sm V)\arrow{r}{\Phi}[swap]{\cong}
&P(\chi(\tr\sm V))\arrow[leftsquigarrow]{r}{\delta}
&B_V(\chi(\tr))\arrow[hookrightarrow]{r}{\beta}
&P(\chi(\tr))
\end{tikzcd}
\end{equation}
It follows by a~simple diagram chase that the commutativities in the
diagram on Figure~\ref{cd:mt2} which we have shown imply the equality
of these two maps. This is true for all $S$, such that
$S\subseteq\act\tr$ and $S\cap V=\es$. On the other hand, the
subcomplexes $X_{S,\es,V}(\tr)$, where $S\subseteq\act\tr$, $S\cap
V=\es$, cover $B_V(\tr)$.  As a~matter of fact, the simplicial
isomorphisms $\psi$ and $\delta_V(\tr)$ show that they induce
a~stratification which is isomorphic to the stratification of
$P(\tr\sm V)$ by $X_S(\tr\sm V)$. The fact that they cover $B_V(\tr)$
completely implies that the maps \eqref{eq:s1} and \eqref{eq:s2}
remain the same after the first term is skipped, which is the same as
to say that \eqref{cd:b} commutes.  This concludes the proof.  \qed

\end{document}